\documentclass[aps,floatfix,pra,superscriptaddress,reprint,showpacs,10pt,preprintnumbers,longbibliography]{revtex4-1}
\usepackage[utf8]{inputenc}
\usepackage[pdftex]{graphicx}
\usepackage{float}
\usepackage{amssymb}
\usepackage{amsmath}  
\usepackage{dsfont}
\usepackage{array}
\usepackage{bm}
\usepackage{mathrsfs}
\usepackage{pifont}
\usepackage{multirow}
\usepackage{upgreek}
\usepackage[dvipsnames]{xcolor}
\usepackage[pdftex,
            pdftitle={Towards Finding an Optimal Flight Gate Assignment on a Digital Quantum Computer},
            pdfauthor={Yahui Chai, Lena Funcke, Tobias Hartung, Karl Jansen, Stefan Kuehn, Paolo Stornati, Tobias Stollenwerk},
            bookmarks,
            colorlinks,
            linkcolor=myblue,
            citecolor=mymagenta,
            menucolor=black,
            urlcolor=myblue,
            plainpages=false,
            pdfpagelabels,
            hypertexnames=false]{hyperref}
\usepackage{verbatim}
\usepackage{slashed}
\usepackage[braket,qm]{qcircuit}
\usepackage{cleveref}
\usepackage{bm}
\definecolor{mymagenta}{RGB}{200, 0, 100}
\definecolor{myblue}{RGB}{45, 48, 146}

\graphicspath{{figures/}}

\begin{document}
\title{Towards Finding an Optimal Flight Gate Assignment on a Digital Quantum Computer}
\author{Yahui Chai}
\affiliation{Deutsches Elektronen-Synchrotron DESY, Platanenallee 6, 15738 Zeuthen, Germany}

\author{Lena Funcke}
\affiliation{Transdisciplinary Research Area ``Building Blocks of Matter and Fundamental Interactions'' (TRA Matter) and Helmholtz Institute for Radiation and Nuclear Physics (HISKP), University of Bonn, Nußallee 14-16, 53115 Bonn, Germany}
\affiliation{Center for Theoretical Physics, Co-Design Center for Quantum Advantage, and NSF AI Institute for Artificial Intelligence and Fundamental Interactions, Massachusetts Institute of Technology, 77 Massachusetts Avenue, Cambridge, MA 02139, USA}

\author{Tobias Hartung}
\affiliation{Northeastern University - London, Devon House, St Katharine Docks, London, E1W 1LP, United Kingdom}

\author{Karl Jansen}
\affiliation{Deutsches Elektronen-Synchrotron DESY, Platanenallee 6, 15738 Zeuthen, Germany}

\author{Stefan K{\"u}hn}
\affiliation{Deutsches Elektronen-Synchrotron DESY, Platanenallee 6, 15738 Zeuthen, Germany}
\affiliation{Computation-based  Science  and  Technology  Research  Center,The  Cyprus  Institute,  20  Kavafi  Street,  2121  Nicosia,  Cyprus}

\author{Paolo Stornati}
\affiliation{ICFO-Institut de Ciencies Fotoniques, The Barcelona Institute of Science and Technology, Av. Carl Friedrich Gauss 3, 08860 Castelldefels (Barcelona), Spain}

\author{Tobias Stollenwerk}
\affiliation{Institute for Quantum Computing Analytics (PGI-12), Forschungszentrum J\"ulich, Wilhelm-Johnen-Straße, 52428 J\"ulich, Germany} 

\date{\today}
\preprint{MIT-CTP/5484}
\begin{abstract}
We investigate the performance of the variational quantum eigensolver (VQE) for the optimal flight gate assignment problem. This problem is a combinatorial optimization problem that aims at finding an optimal assignment of flights to the gates of an airport, in order to minimize the passenger travel time. To study the problem, we adopt a qubit-efficient binary encoding with a cyclic mapping, which is suitable for a digital quantum computer. Using this encoding in conjunction with the Conditional Value at Risk (CVaR) as an aggregation function, we systematically explore the performance of the approach by classically simulating the CVaR-VQE. Our results indicate that the method allows for finding a good solution with high probability, and the method significantly outperforms the naive VQE approach. We examine the role of entanglement for the performance, and find that ansätze with entangling gates allow for better results than pure product states. Studying the problem for various sizes, our numerical data show that the scaling of the number of cost function calls for obtaining a good solution is not exponential for the regimes we investigate in this work.
\end{abstract}

\maketitle

\section{Introduction}

In recent years, variational quantum algorithms (VQAs)~\cite{peruzzo2014variational,mcclean2016theory,cerezo2021variational} have become increasingly relevant due to substantial progress in quantum hardware development. Such algorithms typically do not require deep quantum circuits that could only be faithfully executed on fully error-corrected quantum computers. Instead, they are amenable to noisy intermediate-scale quantum (NISQ) devices (see, e.g., Refs.~\cite{peruzzo2014variational,Kandala2017,Kokail2018,Hartung2018,Hempel2018,Barkoutsos_2020_CVaR,Atas2021,fga_vqe} for various proof-of-principle demonstrations). While such algorithms are typically heuristics without proven performance guarantees, there are indications that VQAs can outperform classical algorithms for certain computationally hard problems. Besides various applications for quantum simulations, the solution of combinatorial optimization problems are further candidates of widespread applications that can be tackled with VQAs~\cite{farhi2014quantum}.

In order to assess the potential of VQA approaches for real-world applications, it can be useful to investigate applications beyond purely academic problems and to focus on certain industrial use-cases, as they typically exhibit additional complexity. One such example is the flight-gate assignment (FGA) problem~\cite{kim2017airport,stollenwerk2019flight}.
The FGA problem is a quadratic assignment problem~\cite{finke1987quadratic} with additional constraints, as typical for real-world applications. Previous works mainly investigated the solution of the FGA problem~\cite{stollenwerk2019flight} and related problems~\cite{venturelli2016Job,stollenwerk2017quantum,stollenwerk2021agile} with quantum annealers. Here, the constraints are incorporated into an unconstrained cost function by penalty terms.
These approaches have a number of disadvantages, one of which is the typically exponentially small subspace of valid solutions in the entire Hilbert space~\cite{stollenwerk2021agile}. One method for mitigating this issue is to constrain the algorithm to only search the feasible subspace. This idea was originally proposed for quantum annealing~\cite{hen2016quantum} and later adapted to variational algorithms~\cite{hadfield2019quantum,fga_vqe}. The applicability of the latter approach for FGA was investigated by deriving suitable algorithmic primitives for constraint invariance~\cite{stollenwerk2020toward}. Reference~\cite{fga_vqe} implemented a proof-of-principle VQE for the FGA problem using an encoding incorporating some of the constraints on IBM's quantum hardware, thus demonstrating the suitability of problem for digital quantum devices.

In this paper, we systematically assess the performance of VQE for the FGA problem, by numerically studying its performance using the Conditional Value at Risk (CVaR)~\cite{Barkoutsos_2020_CVaR} as an aggregation function. We adopt an encoding that avoids a dominant subspace of invalid solutions, which is similar to the one of Ref.~\cite{fga_vqe}, with the addition of a cyclic mapping. Our study demonstrates that utilizing this encoding, the CVaR-VQE performs significantly better than the naive encoding used in previous works. From classically simulating the CVaR-VQE for various problem sizes up to 18 qubits, our results indicate that the number of cost function calls to obtain a reasonably large contribution of the optimal solution in the final state does not scale exponentially with the problem size. Furthermore, we examine the role of entangling gates in the ansatz. Our results demonstrate that ansätze creating entanglement between qubits show a significantly better performance than circuits preparing only product states.

The paper is organized as follows. In Sec.~\ref{sec:model}, we first introduce the FGA problem, before discussing the one-hot encoding and the binary encoding of the problem. Subsequently, we discuss the CVaR-VQE method and the types of ansätze we use in our simulations in Sec.~\ref{sec:vqe}. Section~\ref{sec:results} shows our numerical results for classically simulating the CVar-VQE for various problem sizes, and a comparison between entangling ansätze and ansätze that only produce product states. Finally, we conclude in Sec.~\ref{sec:results}.

\section{The flight-gate assignment problem and its encoding into quantum states\label{sec:model}}

In this section, we first introduce the FGA problem and then proceed with discussing two ways of encoding the problem into quantum states: the one-hot encoding, which does not incorporate any of the constraints, and a binary encoding that integrates some of the constraints.

\subsection{The flight-gate assignment problem\label{subsec:FGA}}

The FGA problem aims at minimizing the total transit time of passengers in an airport by finding an optimal gate assignment of the flights. 
Although there are multiple scenarios for optimizing the gate assignment of flights at an airport, 
we choose the one where we seek to minimize the total transfer time of passengers at the airport~\cite{kim2017airport}.
In this scenario, we have three kinds of passengers in an airport: arriving passengers, departing passengers, and transfer passengers. The arriving passengers land at the airport with an inbound flight and need to walk from the arrival gate to the baggage claim before leaving the airport. Departing passengers enter the airport through the security checkpoint and leave with an outbound flight. Transfer passengers arrive at the airport with an inbound flight, have to walk to the gate of their connecting flight, and leave with an outbound flight. To model the problem mathematically given a set of flights $F$ and a set of gates $G$, we consider a set of binary decision variables $x_{i \alpha}$ that represent whether a flight $i$ is assigned to a gate $\alpha$ or not:
\begin{align}
    x_{i \alpha} = \begin{cases}
    1, &\text{if flight $i \in F$ is assigned to gate $\alpha \in G$}, \\
    0, &\text{otherwise}.
\end{cases}
\end{align}
Throughout the paper, we refer to gates with Greek indices, to flights with Latin indices, and $x=(x_{i\alpha})\in\{0,1\}^{|F|\times |G|}$ is a binary vector collecting all of the $|F|\times |G|$ decision variables. The total passenger travel time can then be expressed as a function of $x$ and is given by
\begin{equation}\label{total_obj}
    T(x) = T^\text{arr}(x) + T^\text{dep}(x) + T^\text{trans}(x),
\end{equation}
where the three parts arise from the contributions of the different types of passengers.
The time $T^\text{arr/dep}$ represents the total transit time of arriving/departing passengers and is given by the partial sums 
\begin{equation}\label{arr_dep_obj}
    T^\text{arr/dep}(x) = \sum_{i\alpha} n_i^\text{arr/dep} t_{\alpha}^\text{arr/dep} x_{i\alpha},
\end{equation}
where $n_i^\text{arr/dep}$ is the number of passengers arriving/departing with flight $i$, and $t_{\alpha}^\text{arr/dep}$ is the time it takes to walk from/to gate $\alpha$. The total time $T^\text{trans}$ of the transfer passengers is given by the sum of the times $t_{\alpha\beta}$ that it takes to go from gate $\alpha$ to gate $\beta$ for each of the $n_{ij}^\text{trans}$ passengers who transfer from flight $i$ to flight $j$ (or vice versa), given that flight $i$ is assigned to gate $\alpha$ and flight $j$ is assigned to gate $\beta$,
\begin{equation}\label{trans_obj}
    T^\text{trans}(x) = \sum_{i,j,\alpha,\beta} n_{ij}^\text{trans}t_{\alpha\beta} x_{i\alpha} x_{j\beta}.
\end{equation}
Note that $T^\text{trans}(x)$ contains a term quadratic in the decision variables. Thus, minimizing the total time in Eq.~\eqref{total_obj} is an instance of a quadratic assignment problem, which are in general NP-hard~\cite{Garey1979}.

In addition, there are two constraints in the FGA problem. Firstly, each flight can only be assigned to one gate, so there can only be a single non-zero decision variable among those belonging to the same flight. This constraint can be enforced by imposing
\begin{align}
    \forall i\in F\quad \sum_{\alpha} x_{i\alpha} = 1.
    \label{eq:constraint1}
\end{align}
Secondly, there can be at most a single flight at a gate at the same time, because flights departing at the same time from the airport cannot be assigned to the same gate. This can be expressed as 
\begin{equation}
    \forall \alpha\in G \ \text{and}\  \forall (i,j)\in P\quad  x_{i\alpha}\times x_{j\alpha} = 0,
    \label{eq:constraint2}
\end{equation}
where $P$ is the set of forbidden flight pairs,
\begin{equation}
    P = \{ (i,j) \in F\times F : t_i^\text{in} < t_j^\text{in} < t_i^\text{out} + t^\text{buf} \}.
\end{equation}
In the expression above, $t_i^\text{in/out}$ is the time of arrival/departure of flight $i$, and $t^\text{buf}$ is a buffer time between two flights at the same gate. In the following, we refer to an assignment of the decision variables fulfilling the two constraints above as a feasible assignment.

The encoding presented above requires $|G|$ decision variables $x_{i1}\dots x_{i|G|}$  for each flight $i\in F$, which can be interpreted as a bit string. The constraint in Eq.~\eqref{eq:constraint1} then implies that only a single entry in such a bit string can be nonzero. Hence, we call the encoding presented above the one-hot encoding. Since for each flight only $|G|$ assignments of the corresponding decision variables are compliant with the constraint in Eq.~\eqref{eq:constraint1}, the total number of feasible assignments is upper bounded by $|G|^{|F|}$.
    
\subsection{Hamiltonian formulation using the one-hot encoding}\label{sec:Hamiltonian one-hot}

In order to treat the problem on a quantum computer, we have to formulate the problem as a (quantum) Hamiltonian. In order to minimize the objective function $T(x)$ subject to the constraints in Eqs.~\eqref{eq:constraint1} and \eqref{eq:constraint2}, we want to incorporate the constraints in the objective function. To this end, we translate them to positive semidefinite penalty terms whose kernel corresponds to valid solutions fulfilling the constraints. These penalty terms can then simply be added to the objective function with a large positive constant in front, thus ensuring that the global minimum is the optimal solution fulfilling the constraints. 

Equation~\eqref{eq:constraint1} can be represented as a penalty term,
\begin{equation}
    C^\text{one}(x) = \sum_{i}\left(\sum_{\alpha} x_{i\alpha} - 1\right) ^2, 
    \label{eq:cone}
\end{equation}
while the second constraint in Eq.~\eqref{eq:constraint2} can be formulated as 
\begin{equation}
    C^\text{not}(x) = \sum_{(i,j) \in P} \sum_{\alpha} x_{i\alpha} x_{j\alpha}.
\end{equation}
Considering both the objective function and the penalty terms, the total cost function  can be formulated as a Quadratic Unconstrained Binary Optimization (QUBO) problem:
\begin{equation}
\begin{aligned}
    Q(x) &= T(x) + \lambda^\text{one} C^\text{one}(x) + \lambda^\text{not} C^\text{not}(x). \\
            &= c + \sum_{i\alpha} h_{i\alpha}  \times x_{i\alpha} +  \sum_{i\alpha j\beta} J_{i\alpha j\beta}\times x_{i\alpha} x_{j\beta}.
\end{aligned}
\label{eq:QUBO}
\end{equation}
In the equation above, $c$, $h_{i\alpha}$, and $J_{i\alpha j\beta}$ are the coefficients of the corresponding terms, which depend on $t_\alpha^\text{arr/dep}$, $t_{\alpha\beta}$, $n_i^\text{arr/dep}$, and $n_{ij}^\text{trans}$. The explicit formulas of these coefficients are shown in Eq.~\eqref{eq: coefficients of QUBO} of Appendix~\ref{sec: coefficients of Hamiltonian using one-hot encoding}. The parameters $\lambda^\text{one}$ and $\lambda^\text{not}$ are constants that have to be chosen large enough to ensure the solution of the QUBO problem above satisfies the constraints. For practical purposes, the values of these parameters might have to be set carefully to make the optimization procedure efficient~\cite{stollenwerk2019flight}.

In order to solve this problem using a quantum device, the QUBO problem has to be mapped to a Hamiltonian acting on qubits. This can be easily realized by replacing the binary decision variables $x_{i\alpha}$ in $Q(x)$ with the operators $(I-\hat{Z}_{k})/2$, where $I$ is the identity and $\hat{Z}_{k}$ is the Pauli $Z$-matrix acting on the qubit that encodes the decision variable $x_{i\alpha}$. Substituting this transformation into the QUBO problem in Eq.~\eqref{eq:QUBO}, we obtain the (quantum) Hamiltonian
\begin{equation}
\begin{aligned}
    \hat{H} &=  c^{\prime} \hat{I} + \sum_{p}^N h_p^{\prime} \hat{Z}_p + \sum_{p<q}^N J^{\prime}_{pq} \hat{Z}_p \hat{Z}_q,
\end{aligned}
\label{eq:IsingHamiltonian}
\end{equation}
where $N = |F| \times |G|$ and $c^{\prime}$, $h_p^{\prime}$, and $J_{pq}^\prime$ are coefficients related to the ones of the original QUBO problem (see Eq.~\eqref{eq: coefficients of Ising Hamiltonian} in Appendix~\ref{sec: coefficients of Hamiltonian using one-hot encoding} for details). The bit strings $x$ are now encoded by a computational basis state $\ket{x}$, and we call $\ket{x}$ a feasible state if $x$ represents a feasible assignment. The optimal solution of the FGA problem subject to the constraints corresponds to the ground state of the Hamiltonian above. By construction, the ground state will be a computational basis state since $\hat{H}$ is diagonal in the $Z$-basis.

Note that in the encoding presented above, each decision variable is mapped to a single qubit. Hence, a total number of $|F|\times |G|$ qubits are required to address the problem on a quantum computer. However, only $|G|^{|F|}$ of the $2^{|G|\times |F|}$ basis states correspond to an assignment for which Eq.~\eqref{eq:cone} is zero. Hence, the fraction of states in the Hilbert space fulfilling the first constraint, and correspondingly the number of feasible states, will decay exponentially with the problem size:
\begin{equation}\label{feasible_ratio_one}
    R_\text{fea}^\text{one}  =\left( \frac{|G|}{2^{|G|}} \right)^{|F|}.
\end{equation} 
As a result, searching for the optimal solution will become increasingly challenging for increasing numbers of flights and gates. 
    
\subsection{Hamiltonian formulation using a binary encoding}
In order to avoid this exponential decay of the feasible subspace, we use a binary encoding for the FGA problem and derive the corresponding Hamiltonian, which is similar to the efficient embedding in Ref.~\cite{fga_vqe}. In addition, we use a more efficient cyclic mapping as shown below. 

As we have discussed in Sec.~\ref{subsec:FGA}, there are $|G|$ assignments compliant with the first constraint in Eq.~\eqref{eq:constraint1} for the decision variables corresponding to each flight. These assignments can be represented with $M = \lceil \log(|G|) \rceil$ (qu)bits using a binary encoding. Since $|G|$ is in general not a power of 2, we choose to map the elements in $G$ to the $2^M$ basis states $\ket{\alpha^{\prime}}$ cyclically as
\begin{equation}
    \ket{\alpha^{\prime}} \leftrightarrow \text{gate}\ \alpha = \alpha^{\prime}\bmod |G| \in G,
    \label{eq:mapping}
\end{equation}
where $\alpha' = 0, \dots ,2^M-1$. In contrast, the previous work in Ref.~\cite{fga_vqe} added a penalty term for the additional states $\{|\alpha^{\prime}\rangle : |G| \leq \alpha^{\prime} < 2^M\}$, in case $G$ is not a power of 2. However, this will lead to an exponential decay with $|F|$ for the fraction of feasible states, as these are given by $\left(G/2^M\right)^{|F|}$. The cyclic mapping used in this work can avoid this exponential decay of feasible states and will usually lead to many degenerate ground states, rendering it easier to find an optimal solution. All in all, for a total of $|F|$ flights, this encoding allows us to represent all possible assignments with  $|F| \times M$ qubits, a lot less than that required for the one-hot encoding. Moreover, by construction, all solutions in this encoding automatically fulfill the constraint in Eq.~\eqref{eq:constraint1}.

In order to be able to solve the problem on a quantum computer using a VQA, we have to translate the Hamiltonian in Eq.~\eqref{eq:IsingHamiltonian} to this encoding. To this end, we define a set of projection operators $P_i(\alpha^{\prime})$, $i = 0,\cdots ,F-1$ given by
\begin{equation}
    \begin{aligned}
        \hat{P}_i(\alpha^{\prime}) &= \op{\alpha^{\prime}}{\alpha^{\prime}}_i = \op{z_0 \cdots z_{M-1}}{z_0 \cdots z_{M-1}}_i \\
        &= \left( \op{z_0}{z_0}  \otimes \dots \otimes \op{z_{M-1}}{z_{M-1}} \right)_i. 
    \end{aligned}
\end{equation}
In the expression above, $z_0 \cdots z_{M-1}$ is the bit string for the binary representation of $\alpha^{\prime}$ and the index $i$ indicates the set of qubits related to flight $i$, on which the projection operators are acting on. Applying $\hat{P}_i(\alpha^{\prime})$ to one of the basis states encoding the solutions compliant with the first constraint for flight $i$ results in a $1$, if and only if flight $i$ is assigned to gate $\alpha$, $\hat{P}_i(\alpha^{\prime})\ket{\beta'}_i = \delta_{\alpha'\beta'}$. Using these projection operators, the Hamiltonian can be expressed as 
\begin{equation}
    \hat{H}(\hat{Z}) = \hat{H}^\text{arr} + \hat{H}^\text{dep} + \hat{H}^\text{trans} + \lambda^\text{not} \hat{H}^\text{not}, 
    \label{eq:binaryHamiltonian}
\end{equation}
where the individual terms are given by
\begin{equation}
\begin{aligned}
    \hat{H}^\text{arr/dep} &= \sum_{i} \sum_{\alpha^{\prime} = 0}^{2^M-1} n_i^\text{arr/dep} t_{\alpha}^\text{arr/dep} \hat{P}_i(\alpha^{\prime}), \\
    \hat{H}^\text{trans} &= \sum_{ij} \sum_{\alpha^{\prime} \beta^{\prime} = 0}^{2^M-1} n_{ij}^\text{trans} t_{\alpha\beta}^\text{trans}  \hat{P}_i(\alpha^{\prime}) \hat{P}_j(\beta^{\prime}), \\
    \hat{H}^\text{not} &= \sum_{(i,j)\in P} \sum_{\alpha^{\prime} \beta^{\prime}= 0}^{2^M-1} \delta_{\alpha \beta} \hat{P}_i(\alpha^{\prime}) \hat{P}_j(\beta^{\prime}).
\end{aligned}
\end{equation}
In the expression above, $\alpha$ and $\beta$ refer to the gate indices after applying the mapping from Eq.~\eqref{eq:mapping}. Note that we no longer have to impose the first constraint from Eq.~\eqref{eq:constraint1} with a penalty term anymore, as it is fulfilled by construction. Moreover, the Hamiltonian can be easily decomposed into Pauli operators using the relation
\begin{align}
    \op{z_k}{z_k}_i = \left(\hat{I} + (-1)^{z_k}Z_{i\times M + k}\right)/2,
\end{align}
where we have chosen a linear ordering of the qubits.

\begin{figure}[htbp]
    \centering
    \includegraphics[width = 0.95\columnwidth]{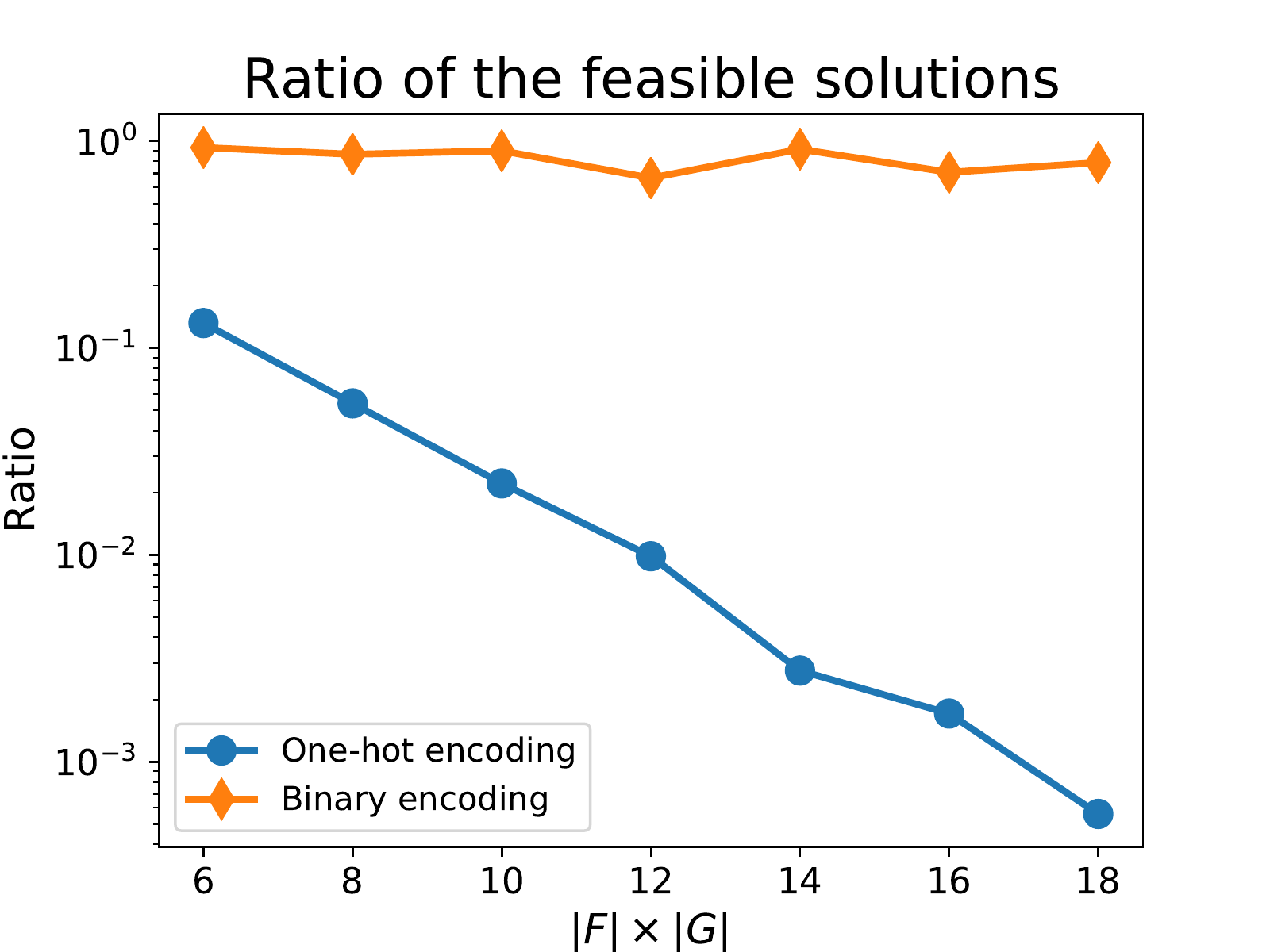}
    \caption{The ratio of the feasible states in the Hilbert space, which are the states that fulfil both the first and the second constraint, as a function of the problem size $|F|\times |G|$ for the one-hot encoding (blue dots) and the binary encoding (orange diamonds). As a guide for the eye, the markers are connected with lines. The ratio of the feasible states for the one-hot encoding decays approximately exponentially with problem size, see Eq.~\eqref{feasible_ratio_one}, which is mainly caused by states violating the first constraint. For details on instance generation, see Sec.~\ref{sec:results}.}
    \label{fig:feasible_ratio}
\end{figure}

The binary encoding with cyclic mapping still allows for unfeasible states, as the second constraint from Eq.~\eqref{eq:constraint2} is not automatically fulfilled. Compared to the exponential decay observed for the one-hot encoding, the ratio of feasible solutions for the binary encoding is a lot larger, and it decays only very slowly with problem size, as shown in Fig.~\ref{fig:feasible_ratio}.
In conjunction with its reduced qubit requirements, the binary encoding with cyclic mapping is significantly more amenable for NISQ devices, which provide only limited resources. The Hamiltonian corresponding to the binary encoding consists of $\mathcal{O}(|F|^2\times |G|^2)$ Pauli $Z$-terms with order $2\times  \lceil \log(G) \rceil$ or less, meaning that each Pauli $Z$-term only acts nontrivially on at most $2\times  \lceil \log(G) \rceil$ qubits. Thus, the expectation value of the Hamiltonian can be evaluated efficiently on a quantum computer.

\section{Variational quantum eigensolver using the conditional value at risk\label{sec:vqe}}

The VQE is a hybrid quantum-classical algorithm for finding an approximation to the ground state of a given Hamiltonian $\hat{H}$ by minimzing $\langle \psi(\bm{\theta})|\hat{H}|\psi(\bm{\theta})\rangle$. Here, $|\psi(\bm{\theta})\rangle$ is a normalized ansatz state, which is parametrized by real numbers $\bm{\theta}$. To find an optimal set of parameters, the VQE utilizes of a feedback loop between a quantum device and a classical computer. The former is used to realize a variational ansatz $\ket{\psi(\bm{\theta})}$ in form of a parametric quantum circuit, and to measure the expectation value of the Hamiltonian. The classical computer is running a minimization algorithm suggesting a new set of parameters $\bm{\theta}'$ based on the measurement outcome of the quantum device. Running the feedback loop until convergence, the parametric circuit encodes an approximation of the ground state of the given Hamiltonian. Due to its modest quantum hardware requirements, and its partial resilience to noise, the VQE is one of the most promising candidates for applications on NISQ devices. While the VQE was originally proposed for finding the ground state of a molecule~\cite{peruzzo2014variational}, it can be readily applied to many other fields~(see, e.g., Refs.~\cite{Kokail2018,Paulson2020,Avkhadiev2020,Mazzola2021,Tilly2021}). 

In particular, the VQE has been proposed to solve combinatorial optimization problems~\cite{Amaro_2022_JSP,Nannicini_2019_IBM,Mugel2022}. Contrary to strongly-correlated quantum many-body systems, for combinatorial optimization problems the problem Hamiltonian is diagonal and the possible solutions correspond to basis states. Since we are only interested in obtaining a good candidate for the solution of the combinatorial optimization problem, the resulting state at the end of the VQE does not necessarily have to be dominated by the state encoding this solution. As long as it produces a state that has a reasonably large component of such a solution, the projective measurements at the end will reveal it, provided enough measurements are taken. Due to this property,  Ref.~\cite{Barkoutsos_2020_CVaR} argued that the CVaR is better suited as a cost function for combinatorial optimization problems than the expectation value of the Hamiltonian. The CVaR for a random variable $X$ with the cumulative density function $F_X$ is defined as the conditional expectation over the left $\xi$-tail of the distribution,
\begin{equation}
    \text{CVaR}_{\xi}(X) = \mathbb{E}\left[ X | X \leq F^{-1}_X(\xi) \right],
\end{equation}
where $\xi \in (0,1]$. This can be applied to VQE by considering only a subset of the samples obtained during the measurement process. Suppose we perform $K$ measurements resulting in the bit strings $\{z_1, z_2, \cdots ,z_K\}$ and the corresponding energy values $\{E_1, E_2, \cdots ,E_K\}$. Assuming the energy values are sorted in ascending order, the CVaR can be calculated as
\begin{equation}
    \text{CVaR}_{\xi} = \frac{1}{\lceil \xi K \rceil} \sum_{i=1}^{\lceil \xi K \rceil} E_i.
    \label{eq:cvar}
\end{equation}
Note that for $\xi=1$ the $\text{CVaR}_{\xi}$ is nothing but the usual estimate for the expectation value with $K$ samples. In the opposite limit, $\xi \to 0$, the $\text{CVaR}_{\xi}$ corresponds to selecting the measurement that produced the lowest energy. Moreover, the definition in Eq.~\eqref{eq:cvar} shows that the $\text{CVaR}_{\xi}$ does essentially not reward increasing the fidelity of the VQE solution with the ground state beyond $\xi$, as we only consider the subset of the $\lceil \xi K \rceil$ measurements with the lowest energy.

In the following, we use VQE with the $\text{CVaR}_{\xi}$ as a cost function to address the FGA problem. In particular, we explore the performance for various choices of $\xi$ as a function of problem size.

\section{Simulation results\label{sec:results}}
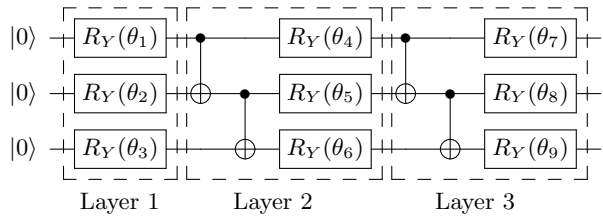
\begin{figure}[htp!]
    \centering
    \begin{align*}
        \Qcircuit @C=1em @R=.7em {
        \lstick{\ket{0}} & \gate{R_Y(\theta_1)} & \ctrl{1}                             & \qw      & \gate{R_Y(\theta_4)} & \ctrl{1}                             & \qw      & \gate{R_Y(\theta_7)} & \qw\\
        \lstick{\ket{0}} & \gate{R_Y(\theta_2)} & \targ                                & \ctrl{1} & \gate{R_Y(\theta_5)} & \targ                                & \ctrl{1} & \gate{R_Y(\theta_8)} & \qw \\
        \lstick{\ket{0}} & \gate{R_Y(\theta_3)} & \qw \gategroup{1}{2}{3}{2}{.9em}{--} & \targ    & \gate{R_Y(\theta_6)} & \qw \gategroup{1}{3}{3}{5}{.9em}{--} & \targ    & \gate{R_Y(\theta_9)} & \qw \gategroup{1}{6}{3}{8}{.9em}{--}\\
        }\\
        \text{Layer 1}\hspace{3em} \text{Layer 2}\hspace{5em} \text{Layer 3}\hspace{3.5em}
    \end{align*}
    \caption{EfficientSU2 ansatz with linear CNOT entangling layers shown for $l=3$ layers and 3 qubits.}
    \label{fig:cnot_circuit}
\end{figure}

In order to explore the performance of the VQE using the CVaR for the FGA problem, we perform classical simulations using the Qiskit~\cite{Qiskit} framework, assuming a perfect quantum device without shot noise, which means we evaluate the cost function exactly. For our experiments, we use the EfficentSU2 ansatz from Qiskit consisting of parametric $R_Y(\theta)=\exp(-i\theta Y / 2)$ rotation gates and linear entangling layers of CNOT gates (see Fig.~\ref{fig:cnot_circuit} for an illustration).
The classical minimization is performed with constrained optimization by linear approximation (COBYLA)~\cite{Powell1994}, whose maximum number of function evaluations is set to 50 times the number of qubits, in order to avoid too long optimization times.

In the following, we examine three aspects. First, we investigate the performance of the VQE using the CVaR for various values of $\xi$ as a function of the problem size. Second, we explore the effect of entanglement on the optimization by using an ansatz that generates product states only, and we compare the results to the ones obtained with the EfficentSU2 ansatz. Finally, we explore the scaling of the method with the problem size.    

\subsection{Performance of the VQE using the CVaR\label{subsec:CVaRSimulationResults}}
       
To investigate the performance of the VQE using the CVaR for the FGA problem, we use the CVaR-VQE to explore various instances of the FGA problem for both the one-hot encoding and the binary encoding with up to 18 qubits. For the one-hot encoding, the number of qubits is equal to the problem size $|F|\times |G| \in \{6,8,10,12,14,16,18\}$. For the binary encoding, we can solve the FGA problem up to $|F|\times |G| = 34$, due its better resource efficiency. For each problem size, we randomly generate a set of non-trivial instances with multiple flights and gates, and pick the ones that are difficult for the classical solver, meaning the ones which take the longest time to solve. We investigate $50$ random instances of the FGA problem for each problem size, and for each instance we run the VQE five times using random choices for the initial parameters in the ansatz. Hence, in total, we explore 250 random instances for each problem size. Moreover, we study the dependence of the results on the choice of the parameter $\xi$ and the number of layers $l$ in the ansatz. To this end, we run simulations with $\xi \in \{0.01, 0.1, 0.25, 1\}$ and $l = 1, 2, 3$.

\begin{figure*}[htbp]
    \centering
    \includegraphics[width=0.9\textwidth]{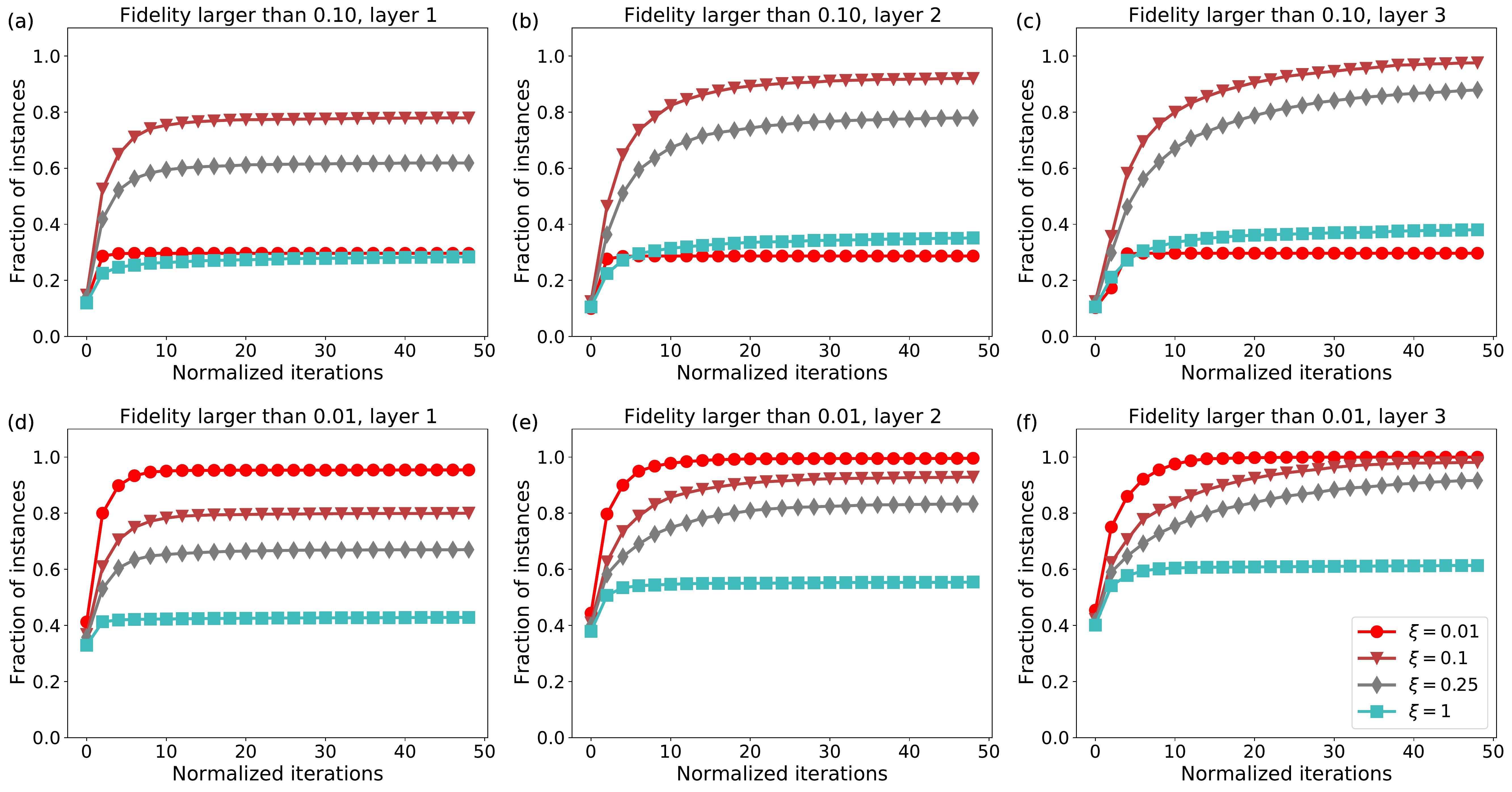}
    \caption{Fraction of instances attaining a fidelity with the exact solution state of at least 10\% (upper row) and 1\% (lower row) as a function of the number of normalized iterations using the binary encoding. Different markers correspond to different choices of $\xi=0.01$ (red dots), 0.1 (dark red triangles), 0.25 (grey diamonds), and 1 (cyan squares). The columns correspond to different numbers of layers $l = 1$ (first column), $2$ (second column) and $3$ (third column).}
    \label{fig:binary iteration}
\end{figure*}

First, we monitor the fraction of instances that reach a certain fidelity threshold 
with the ground state as a function of the iteration number. The fidelity is defined as the probability of sampling a ground state from the quantum state $|\psi\rangle$ prepared by the VQE, $\sum_{z^*}| \langle z^*|\psi\rangle |^2$, where $\{|z^*\rangle \}$ is the set of the ground states that might be degenerate because of the cyclic mapping in Eq.~\eqref{eq:mapping}.
To be able to combine data for different problem sizes, we follow Ref.~\cite{Barkoutsos_2020_CVaR} and consider the normalized number of iterations corresponding to the number of cost function evaluations divided by the number of qubits. Figure~\ref{fig:binary iteration} shows the results for the binary encoding and fidelity thresholds $1\%$ and $10\%$ as a function of the normalized iterations. Comparing the different columns of Fig.~\ref{fig:binary iteration}, corresponding to a different number of layers in the ansatz, we observe that adding more layers yields in general better results. In particular, going beyond a single rotation layer, in which case the ansatz is able to produce entangled states, the fraction of instances that reaches the threshold at the end of the simulation increases noticeably. Moreover, we observe that the CVaR-VQE is able to generate a significant fraction of instances above the fidelity threshold within just a few normalized iterations. The latter indicates that even on NISQ hardware, where one might be restricted to a small number of iterations, one has a reasonable chance of finding a good solution.

Focusing on the results for a fidelity threshold of $1\%$ in Figs.~\ref{fig:binary iteration}(d)-(f), we see that decreasing $\xi$ improves the results. In particular, the conventional VQE using the expectation value of the Hamiltonian as a cost function shows the worst performance, and reaches the $1\%$ fidelity threshold for no more than $60\%$ of all instances, even for three layers (see Fig.~\ref{fig:binary iteration}(f)). Considering a larger fidelity threshold of $10\%$, shown in Figs.~\ref{fig:binary iteration}(a)-(c), the observation is qualitatively similar, except for $\xi=0.01$. The poor performance of $\xi=0.01$ in that case can be explained by the nature of the CVaR cost function. As outlined in Sec.~\ref{sec:vqe}, the CVaR cost function does not reward increasing the fidelity with the ground state beyond $\xi$. Thus, a choice of $\xi=0.01$ does in general not allow for reliably reaching a fidelity with the ground state of $10\%$. Interestingly, the conventional VQE (corresponding to $\xi=1$) only shows a slightly better performance than the CVaR-VQE with $\xi=0.01$. Increasing $\xi$ to the fidelity threshold, we again observe good performance, and for three layers more than $95\%$ of all instances reach a fidelity of at least $10\%$ with the exact solution. Note that the CVaR-VQE also has a better performance in optimizing the QUBO Hamiltonian using one-hot encoding, and achieves a quite high success rate up to 18 variables (qubits) if one use $\xi=0.1$ and three layers (see Fig.~\ref{fig:qubo_iteration} in Appendix~\ref{appendix} for details). 

In our theoretical study, in which we evaluate the cost function exactly, we observe a higher success rate and faster convergence for a lower fidelity threshold and lower values of $\xi$. On quantum hardware, one has to consider that the measurement process involves taking a finite number of samples. Smaller values for $\xi$ imply discarding a larger fraction of samples and, thus, fewer statistics when estimating the cost function using sampling results. Hence, for simulations on quantum hardware, $\xi$ has to be chosen carefully. Reference~\cite{Barkoutsos_2020_CVaR} suggests a choice of $\xi$ in the range of $[0.1, 0.25]$ based on empirical results on quantum hardware. In addition, current NISQ devices suffer from a noticeable level of noise that might affect the results further. In this work, we focus on benchmarking the performance of the VQE on the FGA problem in an ideal setting, and investigating the best choice of $\xi$ for NISQ hardware is beyond the scope of this paper. Hence, for simplicity we focus on two scenarios, $\xi \in \{0.1, 1\}$ with a fidelity threshold of 0.1, as well as $\xi \in \{0.01, 1\}$ with a fidelity threshold of 0.01. 

\begin{figure*}[htbp]
    \centering
    \includegraphics[width=0.9\textwidth]{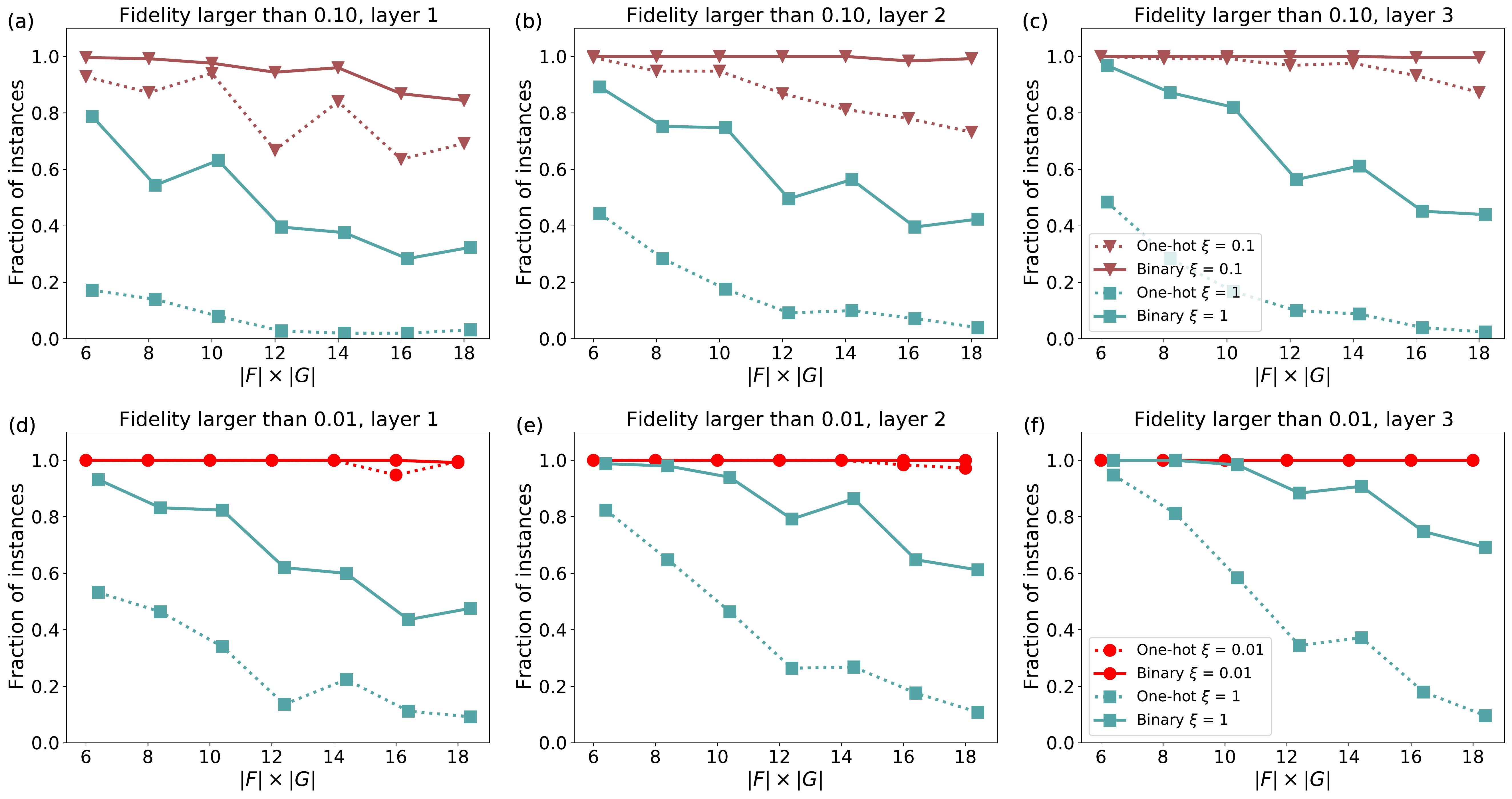}
    \caption{Fraction of instances attaining a fidelity with the exact solution state of at least 10\% (upper row) and 1\% (lower row) at the end of the VQE for problems with up 18 variables, corresponding to up to 18 qubits for the one-hot encoding and up to 9 qubits for the binary encoding. Different markers represent the different choices of $\xi=0.01$ (red dots), 0.1 (brown triangles), and 1 (cyan squares). The columns correspond to different numbers of layers $l = 1$ (first column), $2$ (second column), and $3$ (third column). The results for the binary encoding are connected with solid lines, the ones for the one-hot encoding with dashed lines.}
    \label{fig:QUBO_Binary}
\end{figure*}

Using these scenarios, we compare the performance of both encodings utilizing the CVaR-VQE in Fig.~\ref{fig:QUBO_Binary}. For each problem size, we monitor the fraction of instances whose maximal fidelity throughout the whole optimization process reaches the fidelity threshold. For a fidelity threshold of $10\%$, the binary encoding has a significantly better performance, especially in the case of $\xi=1$, which corresponds to conventional VQE using the expectation value as the cost function (see Figs.~\ref{fig:QUBO_Binary}(a)-(c)). For our largest problem size, $|F|\times |G| = 18$, three layers of the ansatz, and $\xi=1$, the one-hot encoding only reaches the fidelity threshold for a few percent of the instances, while for the binary encoding around $50\%$ of all instances still produce a fidelity of at least $10\%$ (see Fig.~\ref{fig:QUBO_Binary}(c)). Decreasing the value of $\xi$ to 0.1, the drop in the fraction of instances reaching the fidelity threshold with increasing $|F|\times |G|$ is still more pronounced for the one-hot encoding. This reflects the hardness of the VQE using a hardware-efficient ansatz and a normal expectation value as the cost function: it is almost impossible to find the optimal solution of the FGA problem using conventional VQE if the number of qubits is larger than 18. Fortunately, one can use the CVaR as a cost function to overcome this problem, which shows a quite high success rate even for the largest problem size studied in this work. Moreover, while the one-hot encoding can reach the fidelity threshold of $1\%$ with a similarly high success rate as the binary encoding if one chooses $\xi = 0.01$ (see Figs.~\ref{fig:QUBO_Binary}(d)-(f)), the average number of function evaluations to achieve the fidelity $1\%$ is a lot less for the binary encoding compared to the one-hot encoding for the same problem size. As we will examine in detail in Sec.~\ref{scaling of function evaluations}, we observe that the average number of function evaluations for our largest problem size to reach the fidelity threshold of $1\%$ with $\xi = 0.01$ is about $\mathcal{O}(10)$ for the binary encoding and $\mathcal{O}(10^2)$ for the one-hot encoding (see also Fig.~\ref{fig:Binary evaluations}(b) and Fig.~\ref{fig:QUBO_evaluations}(b)).

\subsection{Effect of entanglement on the performance}

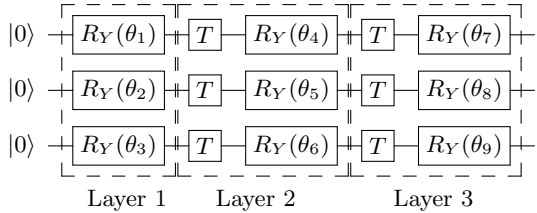
\begin{figure}[htp!]
    \centering
    \begin{align*}
        \Qcircuit @C=1em @R=.7em {
        \lstick{\ket{0}} & \gate{R_Y(\theta_1)}                                  & \gate{T} & \gate{R_Y(\theta_4)}                                  & \gate{T} & \gate{R_Y(\theta_7)} & \qw\\
        \lstick{\ket{0}} & \gate{R_Y(\theta_2)}                                  & \gate{T} & \gate{R_Y(\theta_5)}                                  & \gate{T} & \gate{R_Y(\theta_8)} & \qw \\
        \lstick{\ket{0}} & \gate{R_Y(\theta_3)} \gategroup{1}{2}{3}{2}{.9em}{--} & \gate{T} & \gate{R_Y(\theta_6)} \gategroup{1}{3}{3}{4}{.9em}{--} & \gate{T} & \gate{R_Y(\theta_9)} & \qw \gategroup{1}{5}{3}{6}{.9em}{--}\\
        }\\
        \text{Layer 1}\hspace{2em} \text{Layer 2}\hspace{4em} \text{Layer 3}\hspace{2.5em}
    \end{align*}
    \caption{Quantum circuit without entangling gates, which prepares a product state, illustrated for three qubits.}
    \label{fig:T_circuit}
\end{figure}

The simulation results for both the binary encoding and the one-hot encoding improve when using a larger number of layers in the ansatz, as Fig.~\ref{fig:binary iteration} and Fig.~\ref{fig:QUBO_Binary} (and also Fig.~\ref{fig:qubo_iteration} in Appendix~\ref{appendix}) reveal. However, it is not clear if the improvement of the performance is due to an increased number of entangling layers, or merely because of the presence of more parameters in the ansatz. In order to investigate the role of entanglement in the VQE, we perform simulations using a quantum circuit without entangling gates by replacing the CNOT layers with single-qubit $T$-gates following Ref.~\cite{Nannicini_2019_IBM}. The corresponding circuit is shown in Fig.~\ref{fig:T_circuit}. 

\begin{figure*}[htb]
    \centering
    \includegraphics[width = 0.7\textwidth]{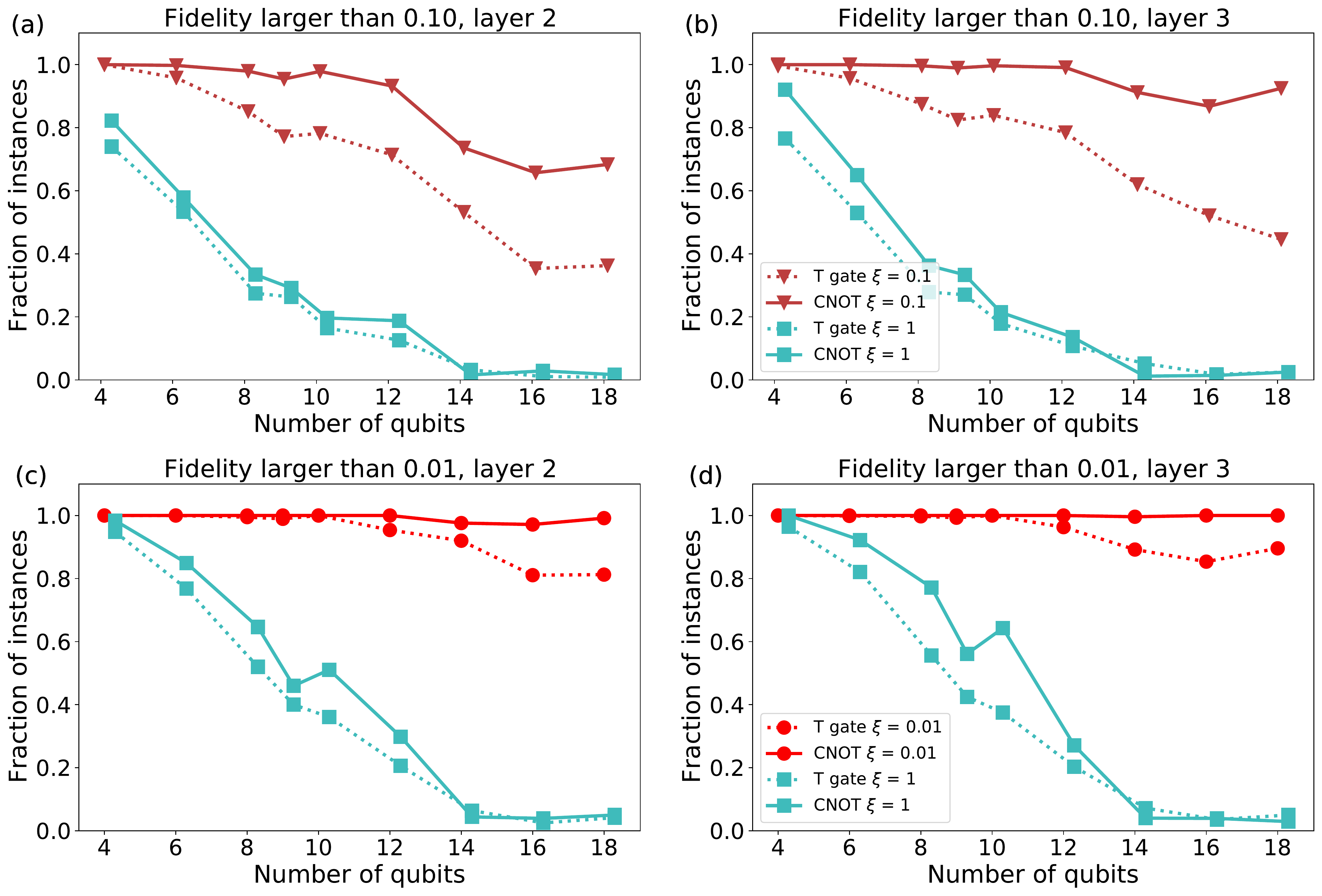}
    \caption{Comparison of the performance of quantum circuits with and without entanglement, for solving the FGA problem using binary encoding.  The circuit with entanglement is the EfficientSU2 circuit with CNOT layers (solid lines); in the circuit without entanglement, the CNOT layers are replaced with single-qubit $T$-gates (dotted lines).  Similar to Fig.~\ref{fig:QUBO_Binary}, the fraction of instances are shown that attain a fidelity with the exact solution state of at least 10\% (upper rown) and 1\% (lower row) at the end of the VQE. The columns correspond to different numbers of layers $l=2$ (first column) and $3$ (second column). Different markers represent different choices of $\xi=0.01$ (red dots), $0.1$ (brown triangles), and $1$ (cyan squares).}
    \label{fig:Binary entanglement}
\end{figure*}

Figure~\ref{fig:Binary entanglement} shows the results for the performance of the CVaR-VQE as a function of the problem size with and without entanglement, using the binary encoding for the FGA problem. 
For two and three layers in the ansatz, the VQE with entanglement shows a clear advantage for both, $\xi = 0.1$ and $\xi = 0.01$, especially for larger problems sizes. In contrast, the advantage of the entangling circuit is not obvious for the VQE using the conventional expectation value, corresponding to $\xi=1$. For the one-hot encoding, we observe a similar behavior; however,  the difference between both circuits is a lot smaller than that for the binary encoding (see Appendix~\ref{appendix}, Fig.~\ref{fig:QUBO_entanglement}). Ideally, this comparison should be extended beyond using only product states, thus also including correlations between qubits that can be efficiently generated by a classical computer, as we aim to address in future work.

\subsection{Scaling of the number of cost function evaluations\label{scaling of function evaluations}}
       
The results shown in the previous sections indicate that CVaR-VQE in conjunction with the binary encoding is suitable to effectively solve the FGA problem. In the following, we examine the scaling of the number of cost function calls with the problem size during the classical minimization, in order to benchmark the efficiency. To this end, we study the average number of cost function evaluations of the successful instances that achieved the chosen fidelity threshold for the final state, which we refer to as $\overline{N}$ in the following. We adopt a best-case scenario and count the number of cost function evaluations until the quantum state generated by the VQE achieves the desired fidelity threshold for the first time~\footnote{Note that this is possible because we assume a perfect quantum computer without shot noise, i.e. we have direct access to the state vector, and we can monitor the overlap with exact solution throughout the optimization procedure.}. 

\begin{figure*}[htbp]
    \centering
    \includegraphics[width = 0.7\textwidth]{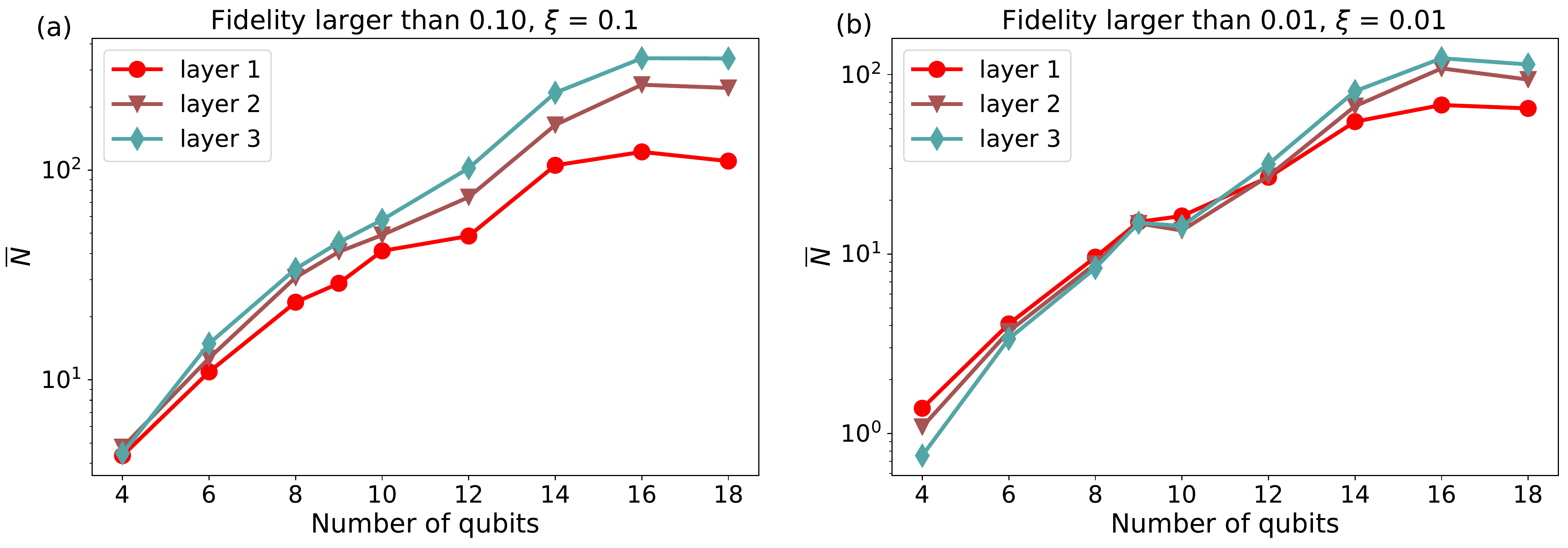}
    \caption{Scaling of the average number of function evaluations $\overline{N}$ with the number of qubits for binary encoding. The simulation results are obtained using $l=1$ (red dots), $2$ (brown triangles) and $3$ (cyan triangles) for the EfficientSU2 circuit. The left panel corresponds to a fidelity threshold of $10\%$ with $\xi=0.1$, the right panel to a threshold of $1\%$ with $\xi=0.01$. The $y$-axes are in logarithmic scale.}
    \label{fig:Binary evaluations}
\end{figure*}

In Fig.~\ref{fig:Binary evaluations}, we show the scaling of $\overline{N}$ for the fidelity thresholds $1\%$ and $10\%$. 
For both cases, we observe a rather similar behavior. After an initially approximately exponential increase of the number of cost function evaluations, the number eventually begins to saturate around a problem size of $14$ qubits. While for a threshold of $10\%$ with $\xi=0.1$, the value of the final plateau in the number of cost function calls shows a slight dependence on the number of layers, this dependence seems to be weaker for a threshold of $1\%$ with $\xi=0.01$. Moreover, a comparison between Fig.~\ref{fig:Binary evaluations}(a) and Fig.~\ref{fig:Binary evaluations}(b) shows that lowering the fidelity threshold and the value of $\xi$ by one order of magnitude only slightly reduces the number of cost function calls required until convergence. Regarding the random instances we used here, some instances have degenerate ground states because of the cyclic mapping in Eq.~\eqref{eq:mapping}. In case of a degeneracy, it can be easier to find a ground state; however, this might influence the scaling of the number of cost function calls to reach a certain fidelity. In particular, we observe a slight  ``staircase shape'' of the data in Fig.~\ref{fig:Binary evaluations}. In order to examine the effect of degeneracies carefully, we also explore the same scaling using a set of random instances with $|G|$ equal to a power of 2, such that there are no degenerate ground states due to the cyclic mapping in Eq.~\eqref{eq:mapping}. For this case, we also do not observe an exponential increase of the number of cost function calls up to the problem size of 18 qubits which we study (see Appendix~\ref{app:scaling} for details). While a larger number of qubits is needed to obtain a solid conclusion for the scaling of $\overline{N}$, which is beyond the scope of this work, our results are promising as they indicate the FGA problem can be efficiently solved on digital quantum computers.

\section{Conclusion and outlook\label{sec:conclusion}}

In this work, we systematically explored the performance of the CVaR-VQE for the FGA problem using a resource-efficient binary encoding in conjunction with a cyclic mapping, which is suitable for digital quantum computers. Compared to the one-hot encoding used in Refs.~\cite{stollenwerk2019flight,stollenwerk2020toward}, the binary encoding with cyclic mapping requires a noticeably smaller number of qubits, and allows for addressing the problem in a resource-efficient manner on NISQ devices. In particular, the fraction of solutions in the Hilbert space that are compatible with the constraints is substantially larger than for the case of the one-hot encoding.

We numerically benchmarked the performance of the CVaR-VQE for the binary encoding, and compared it to the previously used one-hot encoding and conventional VQE. We find that using the CVaR~\cite{Barkoutsos_2020_CVaR} as an aggregation function greatly improves the performance for both encodings compared to conventional VQE. In particular, using a fidelity threshold of at least $10\%$ with the exact solution, the CVaR-VQE is able to reach this threshold for more than $80\%$ of all instances, provided that a large enough number of layers is chosen in the ansatz, compared to less than $40\%$ of all instances using conventional VQE. Moreover, the binary encoding performs significantly better than the one-hot encoding. In particular, the CVaR-VQE approach to the binary encoding does not show a noticeable performance decrease with the problem size for the range of parameters we study, in contrast to the one-hot encoding.  

Comparing the performance of the CVaR-VQE for the binary encoding using the entangling EffiecientSU2 ansatz to a simple ansatz without entangling gates (thus, producing only product states), we find that the presence of entanglement significantly improves the results. In particular, for larger problem sizes, our results indicate that entanglement is beneficial for efficiently exploring the energy landscape. Using a fidelity threshold of $10\%$ and three layers, we observe that the entangling ansatz is able to reach this threshold for roughly $90\%$ of all instances for the largest problem size we study, compared to roughly $40\%$ of all instances for the product state ansatz. 

Focusing on the number of cost function calls to obtain a certain fidelity threshold with the exact ground states, our data suggest that this number does not scale exponentially with the problem size for the range of parameters we study. Consequently, the FGA problem seems to be efficiently addressable with digital quantum computers.

The encoding we used in this work can be readily implemented on gate-based quantum devices, and the number of layers and cost function calls in our study seem within reach on existing quantum hardware. In the future, we plan to investigate the performance of the binary encoding and the CVaR-VQE in a realistic scenario with noise and eventually on a quantum device. 

\acknowledgments
S.K.\ acknowledges financial support from the Cyprus Research and Innovation Foundation under projects
 ``Future-proofing Scientific Applications for the Supercomputers of Tomorrow (FAST)'', contract no.\ COMPLEMENTARY/0916/0048, and ``Quantum Computing for Lattice Gauge Theories (QC4LGT)'', contract no.\ EXCELLENCE/0421/0019.
L.F.\ is partially supported by the U.S.\ Department of Energy, Office of Science, National Quantum Information Science Research Centers,
 Co-design Center for Quantum Advantage (C$^2$QA) under contract number DE-SC0012704, by the DOE QuantiSED Consortium under subcontract number 675352,
 by the National Science Foundation under Cooperative Agreement PHY-2019786
 (The NSF AI Institute for Artificial Intelligence and Fundamental Interactions, \url{http://iaifi.org/}),
 and by the U.S.\ Department of Energy, Office of Science, Office of Nuclear Physics under grant contract numbers DE-SC0011090 and DE-SC0021006. 
 P.S.\ acknowledges support from ERC AdG NOQIA; Ministerio de Ciencia y Innovation Agencia Estatal de Investigaciones (PGC2018-097027-B-I00/10.13039/501100011033, CEX2019-000910-S/10.13039/501100011033, Plan National FIDEUA PID2019-106901GB-I00, FPI, QUANTERA MAQS PCI2019-111828-2, QUANTERA DYNAMITE PCI2022-132919, Proyectos de I+D+I ``Retos Colaboración'' QUSPIN RTC2019-007196-7); MICIIN with funding from European Union NextGenerationEU(PRTR-C17.I1) and by Generalitat de Catalunya; Fundació Cellex; Fundació Mir-Puig; Generalitat de Catalunya (European Social Fund FEDER and CERCA program, AGAUR Grant No. 2021 SGR 01452, QuantumCAT U16-011424, co-funded by ERDF Operational Program of Catalonia 2014-2020); Barcelona Supercomputing Center MareNostrum (FI-2022-1-0042); EU Horizon 2020 FET-OPEN OPTOlogic (Grant No 899794); EU Horizon Europe Program (Grant Agreement 101080086 — NeQST), National Science Centre, Poland (Symfonia Grant No. 2016/20/W/ST4/00314); ICFO Internal ``QuantumGaudi'' project; European Union’s Horizon 2020 research and innovation program under the Marie-Skłodowska-Curie grant agreement No 101029393 (STREDCH) and No 847648 (``La Caixa'' Junior Leaders fellowships ID100010434: LCF/BQ/PI19/11690013, LCF/BQ/PI20/11760031, LCF/BQ/PR20/11770012, LCF/BQ/PR21/11840013). Views and opinions expressed in this work are, however, those of the author(s) only and do not necessarily reflect those of the European Union, European Climate, Infrastructure and Environment Executive Agency (CINEA), nor any other granting authority. Neither the European Union nor any granting authority can be held responsible for them.

\appendix

\section{Details of the Hamiltonian formulation using one-hot encoding}\label{sec: coefficients of Hamiltonian using one-hot encoding}

In Sec.~\ref{sec:Hamiltonian one-hot}, the FGA problem is formulated using the one-hot encoding with the cost function in form of a QUBO problem, as shown in Eq.~\eqref{eq:QUBO}. The corresponding Ising-type Hamiltonian is given in Eq.~\eqref{eq:IsingHamiltonian}. In this appendix, we provide the explicit formula of the coefficients for both formulations of the problem.

Considering the QUBO problem in Eq.~\eqref{eq:QUBO}, the coefficients $c$, $h_{i\alpha}$, and $J_{i\alpha j\beta}$ depend on the number of passengers, $n_i^\text{arr/dep}$ and $n_{ij}^\text{trans}$, and the different times, $t_{\alpha}^\text{arr/dep}$ and $t_{\alpha\beta}$, and read
\begin{equation}\label{eq: coefficients of QUBO}
\begin{aligned}
    c &= |F|\times \lambda^\text{one}, \\
    h_{i\alpha} &= n_i^\text{arr} t_{\alpha}^\text{arr} + n_{i}^\text{dep} t_{\alpha}^\text{dep} - 2\lambda^\text{one}, \\
    J_{i\alpha j\beta} &= n_{ij}^\text{trans} t_{\alpha\beta} + \delta_{ij}\lambda^\text{one} + \delta_{\alpha\beta}\delta_{ij}^P\lambda^\text{one},
\end{aligned}
\end{equation}
where $\delta_{ij}^P$ is nonzero if and only if the flights $i$ and $j$ are a forbidden pair of flights,
\begin{align}
    \delta_{ij}^P = \begin{cases}
    1, &\text{if $(i,j) \in P$}, \\
    0, &\text{otherwise}.
\end{cases}
\end{align}
Given a QUBO problem as in Eq.~\eqref{eq:QUBO}, one can easily get the corresponding Ising Hamiltonian by replacing the binary variables $x_{i\alpha}$ in the QUBO with the operators $(I-\hat{Z}_{i\alpha})/2$, 
\begin{equation}\label{eq: coefficients of Ising Hamiltonian}
\begin{aligned}
H &= c + \sum_{i\alpha} h_{i\alpha} \frac{(I-\hat{Z}_{i\alpha})}{2} + \sum_{i\alpha j\beta} J_{i\alpha j\beta} \frac{(I-\hat{Z}_{i\alpha})(I-\hat{Z}_{j\beta})}{4} \\
&=c + \frac{1}{2}\sum_{i\alpha}h_{i\alpha} + \frac{1}{4}\sum_{i\alpha j\beta} J_{i\alpha j\beta} \\
&+ \sum_{i\alpha} \left[ -\frac{1}{2}h_{i\alpha} - \frac{1}{4}\sum_{j\beta}(J_{i\alpha j\beta} + J_{j\beta i\alpha}) + \frac{1}{4}J_{i\alpha i\alpha}\right] \hat{Z}_{i\alpha} \\
&+ \frac{1}{4} \sum_{i\alpha \neq j\beta}J_{i\alpha j\beta} \hat{Z}_{i\alpha}\hat{Z}_{j\beta}.
\end{aligned}
\end{equation}
The index $i\alpha$ can be mapped to the qubit index $p$ by a linear mapping, $p = i \times |G| + \alpha$, which allows for expressing the Hamiltonian as
\begin{equation}
\begin{aligned}
    \hat{H} &=  c^{\prime} \hat{I} + \sum_{p}^N h_p^{\prime} \hat{Z}_p + \sum_{p<q}^N J^{\prime}_{pq} \hat{Z}_p \hat{Z}_q,
\end{aligned}
\end{equation}
with the coefficients
\begin{equation}
\begin{aligned}
    c^{\prime} &= c + \frac{1}{2}\sum_{i\alpha}h_{i\alpha} + \frac{1}{4}\sum_{i\alpha j\beta} J_{i\alpha j\beta}, \\
    h^{\prime}_p &= -\frac{1}{2}h_{i\alpha} - \frac{1}{4}\sum_{j\beta}(J_{i\alpha j\beta} + J_{j\beta i\alpha}) + \frac{1}{4}J_{i\alpha i\alpha}, \\
    J^{\prime}_{pq} &= \frac{1}{4}\left( J_{i\alpha j\beta} + J_{j\beta i\alpha} \right).
\end{aligned}
\end{equation}
In the expression above, the indices $p$ and $q$ correspond to $p = i \times |G| + \alpha$ and $q = j \times |G| + \beta$.

\section{Simulation results of one-hot encoding}\label{appendix}

\begin{figure*}[t]
    \centering
    \includegraphics[width = 0.9\textwidth]{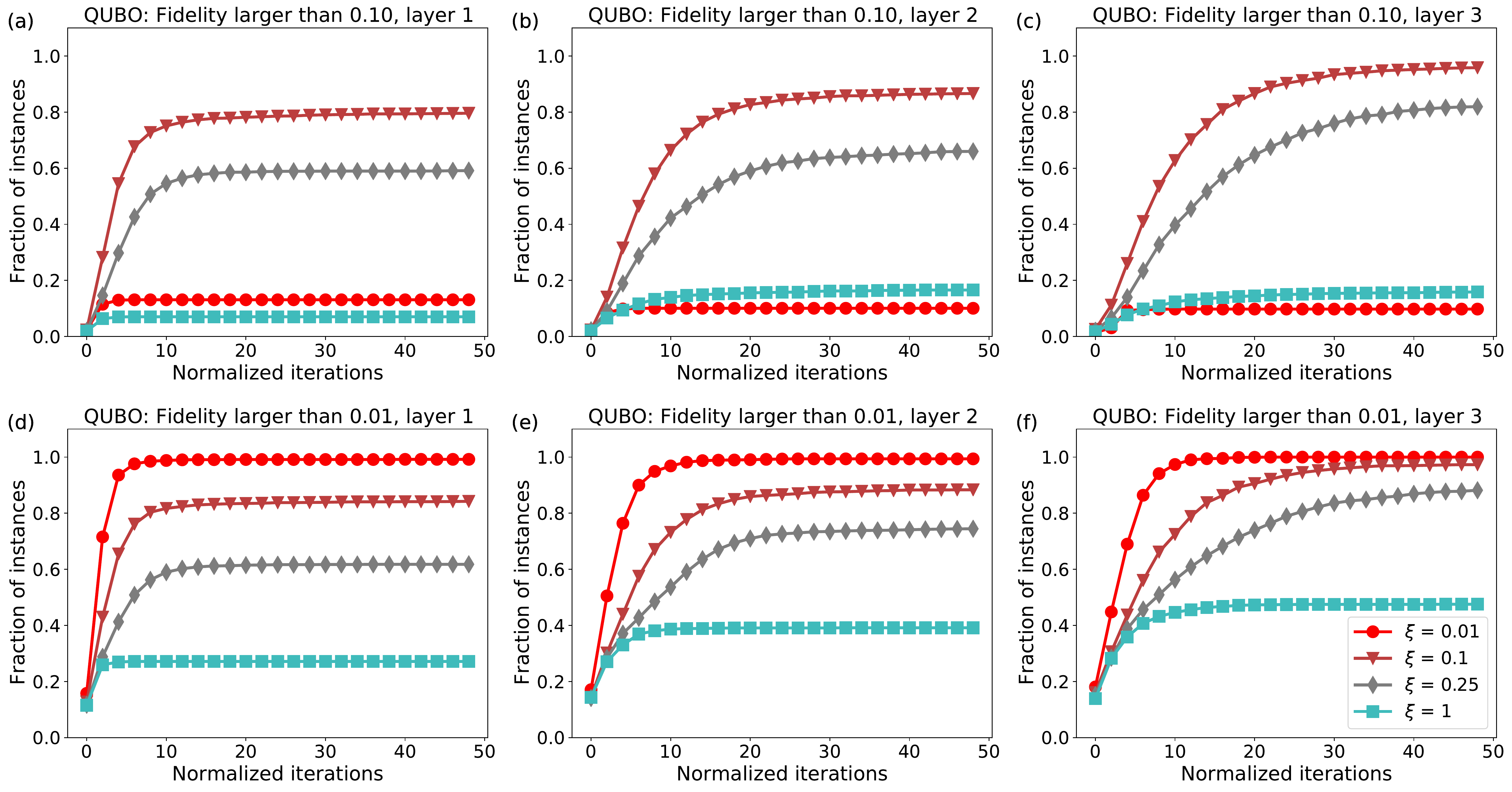}
    \caption{Fraction of instances with the number of normalized iterations using one-hot encoding. There are four different CVaR coefficients $\xi \in \{0.01,0.1,0.25,1\}$ and three different quantum layers: $l \in \{1,2,3\}$ are studied in the plots. As mentioned at Sec.~\ref{subsec:CVaRSimulationResults}, the problem size of the FGA problem in one-hot encoding is from 6 variables to 18 variables, and there are 250 random instances for each problem size, so there are totally 1750 instances in the one-hot encoding for each $\xi$ and each $l$. The plots above show the fraction of instances that achieve the certain fidelity in the 1750 total instances.}
    \label{fig:qubo_iteration}
\end{figure*}

In this appendix, we provide extended simulation results for the FGA problem using the one-hot encoding, which results in the Ising Hamiltonian in Eq.~\eqref{eq:IsingHamiltonian}. As shown in Fig.~\ref{fig:qubo_iteration}, the CVaR-VQE helps to improve the performance and to have a larger number of instances that reach the desired fidelity threshold. In particular, similar to the binary encoding, we observe that a smaller $\xi$ leads to better results. 
    
\begin{figure*}[t]
    \centering    
    \includegraphics[width=0.7\textwidth]{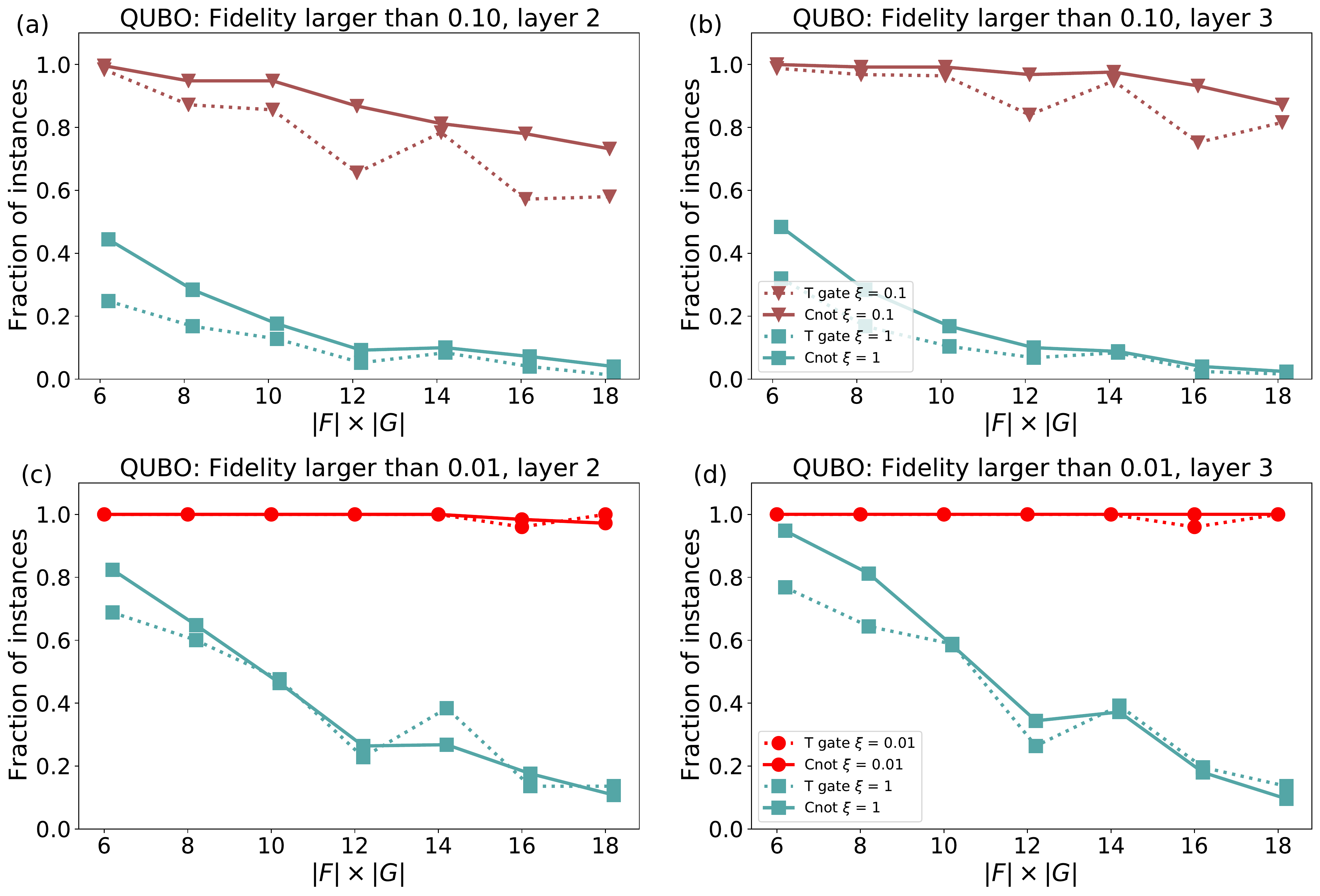}
    \caption{Performance comparison of the quantum circuit with and without entanglement for one-hot encoding. The dotted lines are related to the results obtained by the quantum circuit that only has the one-qubit $R_y$ and T gates (Fig.~\ref{fig:T_circuit}), and the solid lines are related to the circuit with entanglement generated by the CNOT gates (Fig.~\ref{fig:cnot_circuit}). The circuit with entanglement performs better in most cases, especially for the CVaR-VQE with the coefficient $\xi=0.1$.}
    \label{fig:QUBO_entanglement}
\end{figure*}

Figure~\ref{fig:QUBO_entanglement} shows the performance of the one-hot encoding using ansatz circuits with and without entanglement. As the figure reveals, entanglement also plays a positive role in this case; however, the advantage of the entangling circuit is not as great as in the case of the binary encoding, as a comparison with Fig.~\ref{fig:Binary entanglement} reveals. 

\begin{figure*}[t]
    \centering
    \includegraphics[width=0.7\textwidth]{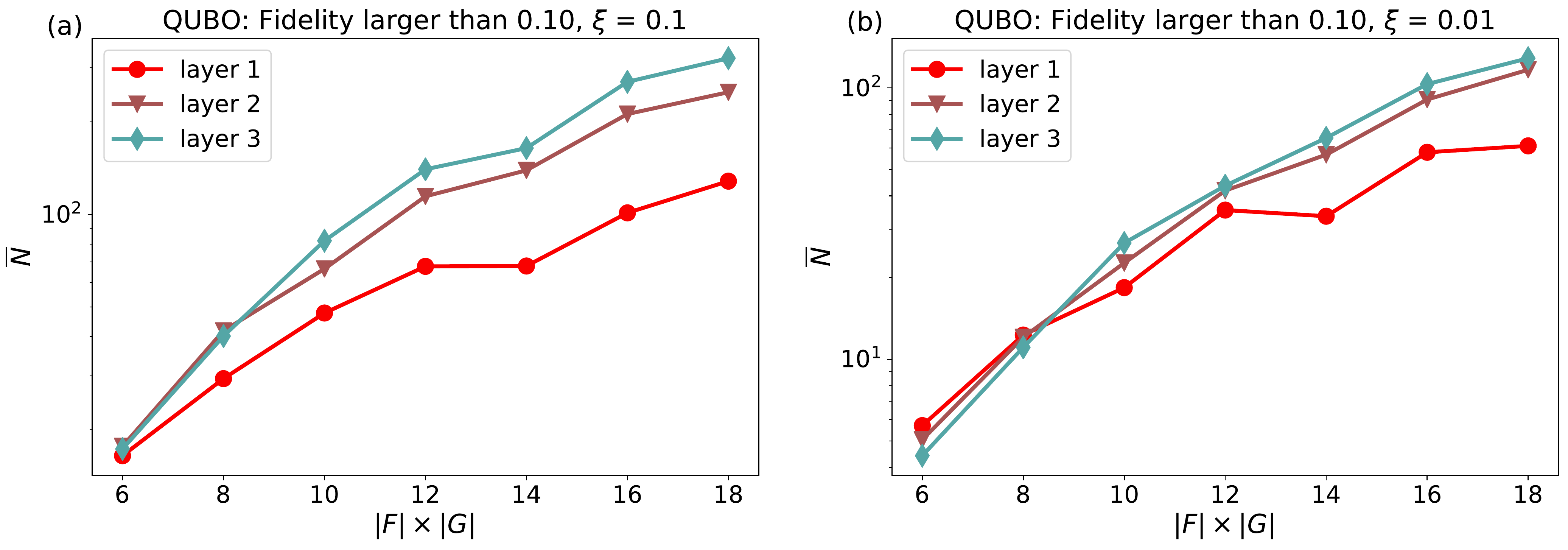}
    \caption{Scaling of the number of the function evaluations ($\overline{N}$) to get the fidelity 0.1 (left) or 0.01 (right) for the FGA problem in the one-hot encoding. The results are obtained using the entangling quantum circuit with the number of layers $l \in \{1,2,3\}$. Just as with the binary encoding (see Sec.~\ref{scaling of function evaluations}), we count the average optimal number of function evaluations of the successful instances that got a fidelity larger than 0.1 or 0.01, respectively.}
    \label{fig:QUBO_evaluations}    
\end{figure*}

Finally, we also explore the scaling of the number of function evaluations of the CVaR-VQE to achieve a certain fidelity threshold in the one-hot encoding. The results for this case can be found in Fig.~\ref{fig:QUBO_evaluations}. For the one-hot encoding, our scaling results are unfortunately inconclusive, as the figure shows. After an initial exponential increase, similar to the binary encoding, it seems that the curve starts to flatten and goes towards a plateau. However, the system sizes we can reach in our classical simulations are too small to solidify this conjecture.

\section{Further exploration of the scaling of the number of cost function evaluations\label{app:scaling}}

\begin{figure*}[hbp!]
    \centering    
    \includegraphics[width=0.7\textwidth]{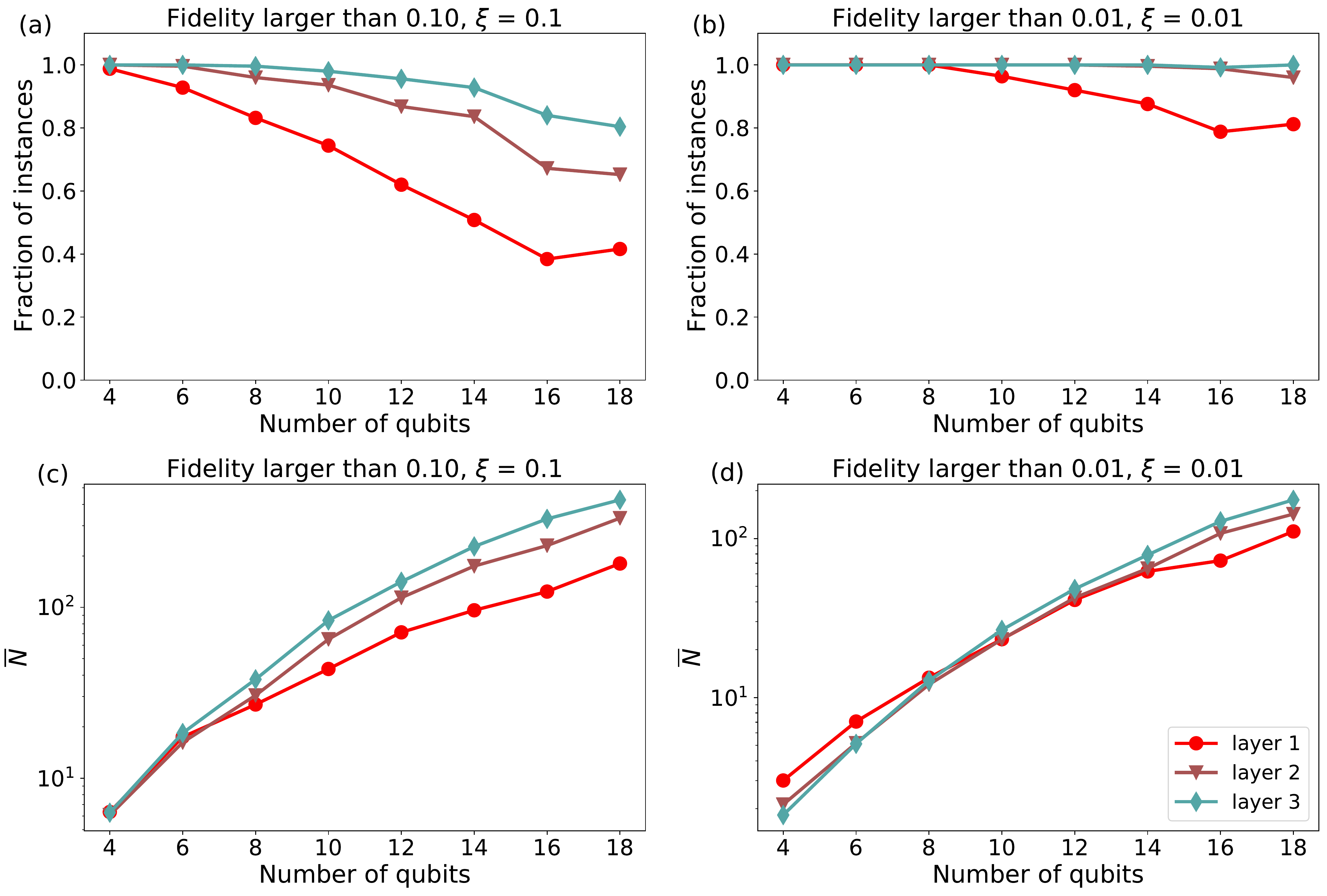}
    \caption{Results of the FGA instances without the degenerate instances in the binary encoding using CVaR-VQE. The upper row shows the fraction of instances that achieve the fidelity threshold using CVaR-VQE with $l=1$ (red dots), $2$ (brown triangles) and 3 (cyan triangles). The results are obtained using a CVaR coefficient of $\xi=0.1$ for the fidelity threshold of $10\%$ and $\xi=0.01$ for the fidelity threshold of $1\%$. The lower row shows the average number of cost function calls ($\overline{N}$) to achieve the fidelity threshold with different quantum layers. The $y$-axes in the panels (c) and (d) are in logarithmic scale.}
    \label{fig:Data_M2_scaling}
\end{figure*}

In this appendix, we examine the scaling of the number of cost function calls with instances that have four gates, $|G|=4$, which can be exactly represented by $M=2$ qubits and will not have degenerate ground states caused by the cyclic mapping in Eq.~\eqref{eq:mapping}. Therefore, we generate a set of random instances with the number of qubits $|F| \times 2 \in \{4,6,8,10,12,14,16,18\}$, with 50 random instances for each problem size and study the CVaR-VQE for five randomly chosen initial parameter sets for each instance. As shown in Fig.~\ref{fig:Data_M2_scaling}(a) and Fig.~\ref{fig:Data_M2_scaling}(b), the fraction of instances that achieve a certain fidelity threshold using the relevant CVaR coefficient ($\xi=0.1$ for fidelity 0.10, $\xi=0.01$ for fidelity 0.01) is still very high, and almost all of instances up to 18 qubits can achieve the fidelity threshold of 0.01 with $\text{CVaR}_{0.01}$. 

Next, we examine the average number of function calls of the instances that achieve the fidelity threshold, which is what we did in Sec.~\ref{scaling of function evaluations} but for different instances. In Fig.~\ref{fig:Data_M2_scaling}(c) and Fig.~\ref{fig:Data_M2_scaling}(d), similar to the scaling of the number of function calls shown above, the curve seems to bend and start to flatten, so we do not observe an exponential scaling up to 18 qubits. However, more qubits are required to get a solid conclusion about the scaling of the number of function calls.

\clearpage
\bibliography{references}

\begin{thebibliography}{31}%
\makeatletter
\providecommand \@ifxundefined [1]{%
 \@ifx{#1\undefined}
}%
\providecommand \@ifnum [1]{%
 \ifnum #1\expandafter \@firstoftwo
 \else \expandafter \@secondoftwo
 \fi
}%
\providecommand \@ifx [1]{%
 \ifx #1\expandafter \@firstoftwo
 \else \expandafter \@secondoftwo
 \fi
}%
\providecommand \natexlab [1]{#1}%
\providecommand \enquote  [1]{``#1''}%
\providecommand \bibnamefont  [1]{#1}%
\providecommand \bibfnamefont [1]{#1}%
\providecommand \citenamefont [1]{#1}%
\providecommand \href@noop [0]{\@secondoftwo}%
\providecommand \href [0]{\begingroup \@sanitize@url \@href}%
\providecommand \@href[1]{\@@startlink{#1}\@@href}%
\providecommand \@@href[1]{\endgroup#1\@@endlink}%
\providecommand \@sanitize@url [0]{\catcode `\\12\catcode `\$12\catcode
  `\&12\catcode `\#12\catcode `\^12\catcode `\_12\catcode `\%12\relax}%
\providecommand \@@startlink[1]{}%
\providecommand \@@endlink[0]{}%
\providecommand \url  [0]{\begingroup\@sanitize@url \@url }%
\providecommand \@url [1]{\endgroup\@href {#1}{\urlprefix }}%
\providecommand \urlprefix  [0]{URL }%
\providecommand \Eprint [0]{\href }%
\providecommand \doibase [0]{http://dx.doi.org/}%
\providecommand \selectlanguage [0]{\@gobble}%
\providecommand \bibinfo  [0]{\@secondoftwo}%
\providecommand \bibfield  [0]{\@secondoftwo}%
\providecommand \translation [1]{[#1]}%
\providecommand \BibitemOpen [0]{}%
\providecommand \bibitemStop [0]{}%
\providecommand \bibitemNoStop [0]{.\EOS\space}%
\providecommand \EOS [0]{\spacefactor3000\relax}%
\providecommand \BibitemShut  [1]{\csname bibitem#1\endcsname}%
\let\auto@bib@innerbib\@empty
\bibitem [{\citenamefont {Peruzzo}\ \emph {et~al.}(2014)\citenamefont
  {Peruzzo}, \citenamefont {McClean}, \citenamefont {Shadbolt}, \citenamefont
  {Yung}, \citenamefont {Zhou}, \citenamefont {Love}, \citenamefont
  {Aspuru-Guzik},\ and\ \citenamefont {O’brien}}]{peruzzo2014variational}%
  \BibitemOpen
  \bibfield  {author} {\bibinfo {author} {\bibfnamefont {Alberto}\ \bibnamefont
  {Peruzzo}}, \bibinfo {author} {\bibfnamefont {Jarrod}\ \bibnamefont
  {McClean}}, \bibinfo {author} {\bibfnamefont {Peter}\ \bibnamefont
  {Shadbolt}}, \bibinfo {author} {\bibfnamefont {Man-Hong}\ \bibnamefont
  {Yung}}, \bibinfo {author} {\bibfnamefont {Xiao-Qi}\ \bibnamefont {Zhou}},
  \bibinfo {author} {\bibfnamefont {Peter~J}\ \bibnamefont {Love}}, \bibinfo
  {author} {\bibfnamefont {Al{\'a}n}\ \bibnamefont {Aspuru-Guzik}}, \ and\
  \bibinfo {author} {\bibfnamefont {Jeremy~L}\ \bibnamefont {O’brien}},\
  }\bibfield  {title} {\enquote {\bibinfo {title} {A variational eigenvalue
  solver on a photonic quantum processor},}\ }\href {\doibase
  10.1038/ncomms5213} {\bibfield  {journal} {\bibinfo  {journal} {Nat.
  Commun.}\ }\textbf {\bibinfo {volume} {5}},\ \bibinfo {pages} {1--7}
  (\bibinfo {year} {2014})}\BibitemShut {NoStop}%
\bibitem [{\citenamefont {McClean}\ \emph {et~al.}(2016)\citenamefont
  {McClean}, \citenamefont {Romero}, \citenamefont {Babbush},\ and\
  \citenamefont {Aspuru-Guzik}}]{mcclean2016theory}%
  \BibitemOpen
  \bibfield  {author} {\bibinfo {author} {\bibfnamefont {Jarrod~R}\
  \bibnamefont {McClean}}, \bibinfo {author} {\bibfnamefont {Jonathan}\
  \bibnamefont {Romero}}, \bibinfo {author} {\bibfnamefont {Ryan}\ \bibnamefont
  {Babbush}}, \ and\ \bibinfo {author} {\bibfnamefont {Al{\'{a}}n}\
  \bibnamefont {Aspuru-Guzik}},\ }\bibfield  {title} {\enquote {\bibinfo
  {title} {The theory of variational hybrid quantum-classical algorithms},}\
  }\href {\doibase 10.1088/1367-2630/18/2/023023} {\bibfield  {journal}
  {\bibinfo  {journal} {New J. Phys.}\ }\textbf {\bibinfo {volume} {18}},\
  \bibinfo {pages} {023023} (\bibinfo {year} {2016})}\BibitemShut {NoStop}%
\bibitem [{\citenamefont {Cerezo}\ \emph {et~al.}(2021)\citenamefont {Cerezo},
  \citenamefont {Arrasmith}, \citenamefont {Babbush}, \citenamefont {Benjamin},
  \citenamefont {Endo}, \citenamefont {Fujii}, \citenamefont {McClean},
  \citenamefont {Mitarai}, \citenamefont {Yuan}, \citenamefont {Cincio} \emph
  {et~al.}}]{cerezo2021variational}%
  \BibitemOpen
  \bibfield  {author} {\bibinfo {author} {\bibfnamefont {Marco}\ \bibnamefont
  {Cerezo}}, \bibinfo {author} {\bibfnamefont {Andrew}\ \bibnamefont
  {Arrasmith}}, \bibinfo {author} {\bibfnamefont {Ryan}\ \bibnamefont
  {Babbush}}, \bibinfo {author} {\bibfnamefont {Simon~C}\ \bibnamefont
  {Benjamin}}, \bibinfo {author} {\bibfnamefont {Suguru}\ \bibnamefont {Endo}},
  \bibinfo {author} {\bibfnamefont {Keisuke}\ \bibnamefont {Fujii}}, \bibinfo
  {author} {\bibfnamefont {Jarrod~R}\ \bibnamefont {McClean}}, \bibinfo
  {author} {\bibfnamefont {Kosuke}\ \bibnamefont {Mitarai}}, \bibinfo {author}
  {\bibfnamefont {Xiao}\ \bibnamefont {Yuan}}, \bibinfo {author} {\bibfnamefont
  {Lukasz}\ \bibnamefont {Cincio}},  \emph {et~al.},\ }\bibfield  {title}
  {\enquote {\bibinfo {title} {Variational quantum algorithms},}\ }\href
  {\doibase 10.1038/s42254-021-00348-9} {\bibfield  {journal} {\bibinfo
  {journal} {Nat. Rev. Phys.}\ }\textbf {\bibinfo {volume} {3}},\ \bibinfo
  {pages} {625--644} (\bibinfo {year} {2021})}\BibitemShut {NoStop}%
\bibitem [{\citenamefont {Kandala}\ \emph {et~al.}(2017)\citenamefont
  {Kandala}, \citenamefont {Mezzacapo}, \citenamefont {Temme}, \citenamefont
  {Takita}, \citenamefont {Brink}, \citenamefont {Chow},\ and\ \citenamefont
  {Gambetta}}]{Kandala2017}%
  \BibitemOpen
  \bibfield  {author} {\bibinfo {author} {\bibfnamefont {Abhinav}\ \bibnamefont
  {Kandala}}, \bibinfo {author} {\bibfnamefont {Antonio}\ \bibnamefont
  {Mezzacapo}}, \bibinfo {author} {\bibfnamefont {Kristan}\ \bibnamefont
  {Temme}}, \bibinfo {author} {\bibfnamefont {Maika}\ \bibnamefont {Takita}},
  \bibinfo {author} {\bibfnamefont {Markus}\ \bibnamefont {Brink}}, \bibinfo
  {author} {\bibfnamefont {Jerry~M.}\ \bibnamefont {Chow}}, \ and\ \bibinfo
  {author} {\bibfnamefont {Jay~M.}\ \bibnamefont {Gambetta}},\ }\bibfield
  {title} {\enquote {\bibinfo {title} {Hardware-efficient variational quantum
  eigensolver for small molecules and quantum magnets},}\ }\href {\doibase
  10.1038/nature23879} {\bibfield  {journal} {\bibinfo  {journal} {Nature}\
  }\textbf {\bibinfo {volume} {549}},\ \bibinfo {pages} {242--246} (\bibinfo
  {year} {2017})}\BibitemShut {NoStop}%
\bibitem [{\citenamefont {Kokail}\ \emph {et~al.}(2019)\citenamefont {Kokail},
  \citenamefont {Maier}, \citenamefont {van Bijnen}, \citenamefont {Brydges},
  \citenamefont {Joshi}, \citenamefont {Jurcevic}, \citenamefont {Muschik},
  \citenamefont {Silvi}, \citenamefont {Blatt}, \citenamefont {Roos},\ and\
  \citenamefont {Zoller}}]{Kokail2018}%
  \BibitemOpen
  \bibfield  {author} {\bibinfo {author} {\bibfnamefont {Christian}\
  \bibnamefont {Kokail}}, \bibinfo {author} {\bibfnamefont {Christine}\
  \bibnamefont {Maier}}, \bibinfo {author} {\bibfnamefont {Rick}\ \bibnamefont
  {van Bijnen}}, \bibinfo {author} {\bibfnamefont {Tiff}\ \bibnamefont
  {Brydges}}, \bibinfo {author} {\bibfnamefont {Manoj~K.}\ \bibnamefont
  {Joshi}}, \bibinfo {author} {\bibfnamefont {Petar}\ \bibnamefont {Jurcevic}},
  \bibinfo {author} {\bibfnamefont {Christine~A.}\ \bibnamefont {Muschik}},
  \bibinfo {author} {\bibfnamefont {Pietro}\ \bibnamefont {Silvi}}, \bibinfo
  {author} {\bibfnamefont {Rainer}\ \bibnamefont {Blatt}}, \bibinfo {author}
  {\bibfnamefont {Christian~F.}\ \bibnamefont {Roos}}, \ and\ \bibinfo {author}
  {\bibfnamefont {Peter}\ \bibnamefont {Zoller}},\ }\bibfield  {title}
  {\enquote {\bibinfo {title} {Self-verifying variational quantum simulation of
  the lattice schwinger model},}\ }\href {\doibase 10.1038/s41586-019-1177-4}
  {\bibfield  {journal} {\bibinfo  {journal} {Nature}\ }\textbf {\bibinfo
  {volume} {569}},\ \bibinfo {pages} {355} (\bibinfo {year}
  {2019})}\BibitemShut {NoStop}%
\bibitem [{\citenamefont {Hartung}\ and\ \citenamefont
  {Jansen}(2019)}]{Hartung2018}%
  \BibitemOpen
  \bibfield  {author} {\bibinfo {author} {\bibfnamefont {Tobias}\ \bibnamefont
  {Hartung}}\ and\ \bibinfo {author} {\bibfnamefont {Karl}\ \bibnamefont
  {Jansen}},\ }\bibfield  {title} {\enquote {\bibinfo {title} {Zeta-regularized
  vacuum expectation values},}\ }\href {\doibase 10.1063/1.5085866} {\bibfield
  {journal} {\bibinfo  {journal} {J. Math. Phys.}\ }\textbf {\bibinfo {volume}
  {60}},\ \bibinfo {pages} {093504} (\bibinfo {year} {2019})}\BibitemShut
  {NoStop}%
\bibitem [{\citenamefont {Hempel}\ \emph {et~al.}(2018)\citenamefont {Hempel},
  \citenamefont {Maier}, \citenamefont {Romero}, \citenamefont {McClean},
  \citenamefont {Monz}, \citenamefont {Shen}, \citenamefont {Jurcevic},
  \citenamefont {Lanyon}, \citenamefont {Love}, \citenamefont {Babbush},
  \citenamefont {Aspuru-Guzik}, \citenamefont {Blatt},\ and\ \citenamefont
  {Roos}}]{Hempel2018}%
  \BibitemOpen
  \bibfield  {author} {\bibinfo {author} {\bibfnamefont {Cornelius}\
  \bibnamefont {Hempel}}, \bibinfo {author} {\bibfnamefont {Christine}\
  \bibnamefont {Maier}}, \bibinfo {author} {\bibfnamefont {Jonathan}\
  \bibnamefont {Romero}}, \bibinfo {author} {\bibfnamefont {Jarrod}\
  \bibnamefont {McClean}}, \bibinfo {author} {\bibfnamefont {Thomas}\
  \bibnamefont {Monz}}, \bibinfo {author} {\bibfnamefont {Heng}\ \bibnamefont
  {Shen}}, \bibinfo {author} {\bibfnamefont {Petar}\ \bibnamefont {Jurcevic}},
  \bibinfo {author} {\bibfnamefont {Ben~P.}\ \bibnamefont {Lanyon}}, \bibinfo
  {author} {\bibfnamefont {Peter}\ \bibnamefont {Love}}, \bibinfo {author}
  {\bibfnamefont {Ryan}\ \bibnamefont {Babbush}}, \bibinfo {author}
  {\bibfnamefont {Al\'an}\ \bibnamefont {Aspuru-Guzik}}, \bibinfo {author}
  {\bibfnamefont {Rainer}\ \bibnamefont {Blatt}}, \ and\ \bibinfo {author}
  {\bibfnamefont {Christian~F.}\ \bibnamefont {Roos}},\ }\bibfield  {title}
  {\enquote {\bibinfo {title} {Quantum chemistry calculations on a trapped-ion
  quantum simulator},}\ }\href {\doibase 10.1103/PhysRevX.8.031022} {\bibfield
  {journal} {\bibinfo  {journal} {Phys. Rev. X}\ }\textbf {\bibinfo {volume}
  {8}},\ \bibinfo {pages} {031022} (\bibinfo {year} {2018})}\BibitemShut
  {NoStop}%
\bibitem [{\citenamefont {Barkoutsos}\ \emph {et~al.}(2020)\citenamefont
  {Barkoutsos}, \citenamefont {Nannicini}, \citenamefont {Robert},
  \citenamefont {Tavernelli},\ and\ \citenamefont
  {Woerner}}]{Barkoutsos_2020_CVaR}%
  \BibitemOpen
  \bibfield  {author} {\bibinfo {author} {\bibfnamefont {Panagiotis~Kl.}\
  \bibnamefont {Barkoutsos}}, \bibinfo {author} {\bibfnamefont {Giacomo}\
  \bibnamefont {Nannicini}}, \bibinfo {author} {\bibfnamefont {Anton}\
  \bibnamefont {Robert}}, \bibinfo {author} {\bibfnamefont {Ivano}\
  \bibnamefont {Tavernelli}}, \ and\ \bibinfo {author} {\bibfnamefont {Stefan}\
  \bibnamefont {Woerner}},\ }\bibfield  {title} {\enquote {\bibinfo {title}
  {Improving variational quantum optimization using {CVaR}},}\ }\href {\doibase
  10.22331/q-2020-04-20-256} {\bibfield  {journal} {\bibinfo  {journal}
  {Quantum}\ }\textbf {\bibinfo {volume} {4}},\ \bibinfo {pages} {256}
  (\bibinfo {year} {2020})}\BibitemShut {NoStop}%
\bibitem [{\citenamefont {Atas}\ \emph {et~al.}(2021)\citenamefont {Atas},
  \citenamefont {Zhang}, \citenamefont {Lewis}, \citenamefont {Jahanpour},
  \citenamefont {Haase},\ and\ \citenamefont {Muschik}}]{Atas2021}%
  \BibitemOpen
  \bibfield  {author} {\bibinfo {author} {\bibfnamefont {Yasar}\ \bibnamefont
  {Atas}}, \bibinfo {author} {\bibfnamefont {Jinglei}\ \bibnamefont {Zhang}},
  \bibinfo {author} {\bibfnamefont {Randy}\ \bibnamefont {Lewis}}, \bibinfo
  {author} {\bibfnamefont {Amin}\ \bibnamefont {Jahanpour}}, \bibinfo {author}
  {\bibfnamefont {Jan~F.}\ \bibnamefont {Haase}}, \ and\ \bibinfo {author}
  {\bibfnamefont {Christine~A.}\ \bibnamefont {Muschik}},\ }\bibfield  {title}
  {\enquote {\bibinfo {title} {Su(2) hadrons on a quantum computer},}\ }\href
  {\doibase 10.1038/s41467-021-26825-4} {\bibfield  {journal} {\bibinfo
  {journal} {Nat Commun.}\ }\textbf {\bibinfo {volume} {12}},\ \bibinfo {pages}
  {6499} (\bibinfo {year} {2021})}\BibitemShut {NoStop}%
\bibitem [{\citenamefont {Mohammadbagherpoor}\ \emph
  {et~al.}(2021)\citenamefont {Mohammadbagherpoor}, \citenamefont {Dreher},
  \citenamefont {Ibrahim}, \citenamefont {Oh}, \citenamefont {Hall},
  \citenamefont {Stone},\ and\ \citenamefont {Stojkovic}}]{fga_vqe}%
  \BibitemOpen
  \bibfield  {author} {\bibinfo {author} {\bibfnamefont {Hamed}\ \bibnamefont
  {Mohammadbagherpoor}}, \bibinfo {author} {\bibfnamefont {Patrick}\
  \bibnamefont {Dreher}}, \bibinfo {author} {\bibfnamefont {Mohannad}\
  \bibnamefont {Ibrahim}}, \bibinfo {author} {\bibfnamefont {Young-Hyun}\
  \bibnamefont {Oh}}, \bibinfo {author} {\bibfnamefont {James}\ \bibnamefont
  {Hall}}, \bibinfo {author} {\bibfnamefont {Richard~E}\ \bibnamefont {Stone}},
  \ and\ \bibinfo {author} {\bibfnamefont {Mirela}\ \bibnamefont {Stojkovic}},\
  }\bibfield  {title} {\enquote {\bibinfo {title} {Exploring airline
  gate-scheduling optimization using quantum computers},}\ }\href
  {https://arxiv.org/abs/2111.09472} {\bibfield  {journal} {\bibinfo  {journal}
  {arXiv:2111.09472}\ } (\bibinfo {year} {2021})}\BibitemShut {NoStop}%
\bibitem [{\citenamefont {Farhi}\ \emph {et~al.}(2014)\citenamefont {Farhi},
  \citenamefont {Goldstone},\ and\ \citenamefont {Gutmann}}]{farhi2014quantum}%
  \BibitemOpen
  \bibfield  {author} {\bibinfo {author} {\bibfnamefont {Edward}\ \bibnamefont
  {Farhi}}, \bibinfo {author} {\bibfnamefont {Jeffrey}\ \bibnamefont
  {Goldstone}}, \ and\ \bibinfo {author} {\bibfnamefont {Sam}\ \bibnamefont
  {Gutmann}},\ }\bibfield  {title} {\enquote {\bibinfo {title} {A quantum
  approximate optimization algorithm},}\ }\href
  {https://arxiv.org/abs/1411.4028} {\bibfield  {journal} {\bibinfo  {journal}
  {arXiv:1411.4028}\ } (\bibinfo {year} {2014})}\BibitemShut {NoStop}%
\bibitem [{\citenamefont {Kim}\ \emph {et~al.}(2017)\citenamefont {Kim},
  \citenamefont {Feron}, \citenamefont {Clarke}, \citenamefont {Marzuoli},\
  and\ \citenamefont {Delahaye}}]{kim2017airport}%
  \BibitemOpen
  \bibfield  {author} {\bibinfo {author} {\bibfnamefont {Sang~Hyun}\
  \bibnamefont {Kim}}, \bibinfo {author} {\bibfnamefont {Eric}\ \bibnamefont
  {Feron}}, \bibinfo {author} {\bibfnamefont {John-Paul}\ \bibnamefont
  {Clarke}}, \bibinfo {author} {\bibfnamefont {Aude}\ \bibnamefont {Marzuoli}},
  \ and\ \bibinfo {author} {\bibfnamefont {Daniel}\ \bibnamefont {Delahaye}},\
  }\bibfield  {title} {\enquote {\bibinfo {title} {Airport gate scheduling for
  passengers, aircraft, and operations},}\ }\href {\doibase 10.2514/1.D0079}
  {\bibfield  {journal} {\bibinfo  {journal} {Journal of Air Transportation}\
  }\textbf {\bibinfo {volume} {25}},\ \bibinfo {pages} {109--114} (\bibinfo
  {year} {2017})}\BibitemShut {NoStop}%
\bibitem [{\citenamefont {Stollenwerk}\ \emph {et~al.}(2019)\citenamefont
  {Stollenwerk}, \citenamefont {Lobe},\ and\ \citenamefont
  {Jung}}]{stollenwerk2019flight}%
  \BibitemOpen
  \bibfield  {author} {\bibinfo {author} {\bibfnamefont {Tobias}\ \bibnamefont
  {Stollenwerk}}, \bibinfo {author} {\bibfnamefont {Elisabeth}\ \bibnamefont
  {Lobe}}, \ and\ \bibinfo {author} {\bibfnamefont {Martin}\ \bibnamefont
  {Jung}},\ }\bibfield  {title} {\enquote {\bibinfo {title} {Flight gate
  assignment with a quantum annealer},}\ }in\ \href
  {https://doi.org/10.1007/978-3-030-14082-3_9} {\emph {\bibinfo {booktitle}
  {Proceedings of the First International Workshop on Quantum Technology and
  Optimization Problems}}},\ \bibinfo {series and number} {\bibinfo {series}
  {Theoretical Computer Science and General Issues}\ No.~\bibinfo {number}
  {9}}\ (\bibinfo  {publisher} {Springer},\ \bibinfo {address} {Munich,
  Germany},\ \bibinfo {year} {2019})\BibitemShut {NoStop}%
\bibitem [{\citenamefont {Finke}\ \emph {et~al.}(1987)\citenamefont {Finke},
  \citenamefont {Burkard},\ and\ \citenamefont {Rendl}}]{finke1987quadratic}%
  \BibitemOpen
  \bibfield  {author} {\bibinfo {author} {\bibfnamefont {Gerd}\ \bibnamefont
  {Finke}}, \bibinfo {author} {\bibfnamefont {Rainer~E.}\ \bibnamefont
  {Burkard}}, \ and\ \bibinfo {author} {\bibfnamefont {Franz}\ \bibnamefont
  {Rendl}},\ }\bibfield  {title} {\enquote {\bibinfo {title} {Quadratic
  assignment problems},}\ }in\ \href {\doibase
  https://doi.org/10.1016/S0304-0208(08)73232-8} {\emph {\bibinfo {booktitle}
  {Surveys in Combinatorial Optimization}}},\ \bibinfo {series} {North-Holland
  Mathematics Studies}, Vol.\ \bibinfo {volume} {132},\ \bibinfo {editor}
  {edited by\ \bibinfo {editor} {\bibfnamefont {Silvano}\ \bibnamefont
  {Martello}}, \bibinfo {editor} {\bibfnamefont {Gilbert}\ \bibnamefont
  {Laporte}}, \bibinfo {editor} {\bibfnamefont {Michel}\ \bibnamefont
  {Minoux}}, \ and\ \bibinfo {editor} {\bibfnamefont {Celso}\ \bibnamefont
  {Ribeiro}}}\ (\bibinfo  {publisher} {North-Holland},\ \bibinfo {year}
  {1987})\ pp.\ \bibinfo {pages} {61--82}\BibitemShut {NoStop}%
\bibitem [{\citenamefont {Venturelli}\ \emph {et~al.}(2016)\citenamefont
  {Venturelli}, \citenamefont {Marchand},\ and\ \citenamefont
  {Rojo}}]{venturelli2016Job}%
  \BibitemOpen
  \bibfield  {author} {\bibinfo {author} {\bibfnamefont {Davide}\ \bibnamefont
  {Venturelli}}, \bibinfo {author} {\bibfnamefont {Dominic J.~J.}\ \bibnamefont
  {Marchand}}, \ and\ \bibinfo {author} {\bibfnamefont {Galo~Higuera}\
  \bibnamefont {Rojo}},\ }\bibfield  {title} {\enquote {\bibinfo {title} {Job
  shop scheduling solver based on quantum annealing},}\ }\href
  {https://arxiv.org/abs/1506.08479} {\bibfield  {journal} {\bibinfo  {journal}
  {arXiv:1506.08479}\ } (\bibinfo {year} {2016})}\BibitemShut {NoStop}%
\bibitem [{\citenamefont {Stollenwerk}\ \emph
  {et~al.}(2020{\natexlab{a}})\citenamefont {Stollenwerk}, \citenamefont
  {O’Gorman}, \citenamefont {Venturelli}, \citenamefont {Mandrà},
  \citenamefont {Rodionova}, \citenamefont {Ng}, \citenamefont {Sridhar},
  \citenamefont {Rieffel},\ and\ \citenamefont
  {Biswas}}]{stollenwerk2017quantum}%
  \BibitemOpen
  \bibfield  {author} {\bibinfo {author} {\bibfnamefont {Tobias}\ \bibnamefont
  {Stollenwerk}}, \bibinfo {author} {\bibfnamefont {Bryan}\ \bibnamefont
  {O’Gorman}}, \bibinfo {author} {\bibfnamefont {Davide}\ \bibnamefont
  {Venturelli}}, \bibinfo {author} {\bibfnamefont {Salvatore}\ \bibnamefont
  {Mandrà}}, \bibinfo {author} {\bibfnamefont {Olga}\ \bibnamefont
  {Rodionova}}, \bibinfo {author} {\bibfnamefont {Hokkwan}\ \bibnamefont {Ng}},
  \bibinfo {author} {\bibfnamefont {Banavar}\ \bibnamefont {Sridhar}}, \bibinfo
  {author} {\bibfnamefont {Eleanor~Gilbert}\ \bibnamefont {Rieffel}}, \ and\
  \bibinfo {author} {\bibfnamefont {Rupak}\ \bibnamefont {Biswas}},\ }\bibfield
   {title} {\enquote {\bibinfo {title} {Quantum annealing applied to
  de-conflicting optimal trajectories for air traffic management},}\ }\href
  {\doibase 10.1109/TITS.2019.2891235} {\bibfield  {journal} {\bibinfo
  {journal} {IEEE Transactions on Intelligent Transportation Systems}\ }\textbf
  {\bibinfo {volume} {21}},\ \bibinfo {pages} {285--297} (\bibinfo {year}
  {2020}{\natexlab{a}})}\BibitemShut {NoStop}%
\bibitem [{\citenamefont {Stollenwerk}\ \emph {et~al.}(2021)\citenamefont
  {Stollenwerk}, \citenamefont {Michaud}, \citenamefont {Lobe}, \citenamefont
  {Picard}, \citenamefont {Basermann},\ and\ \citenamefont
  {Botter}}]{stollenwerk2021agile}%
  \BibitemOpen
  \bibfield  {author} {\bibinfo {author} {\bibfnamefont {Tobias}\ \bibnamefont
  {Stollenwerk}}, \bibinfo {author} {\bibfnamefont {Vincent}\ \bibnamefont
  {Michaud}}, \bibinfo {author} {\bibfnamefont {Elisabeth}\ \bibnamefont
  {Lobe}}, \bibinfo {author} {\bibfnamefont {Mathieu}\ \bibnamefont {Picard}},
  \bibinfo {author} {\bibfnamefont {Achim}\ \bibnamefont {Basermann}}, \ and\
  \bibinfo {author} {\bibfnamefont {Thierry}\ \bibnamefont {Botter}},\
  }\bibfield  {title} {\enquote {\bibinfo {title} {Agile earth observation
  satellite scheduling with a quantum annealer},}\ }\href {\doibase
  10.1109/TAES.2021.3088490} {\bibfield  {journal} {\bibinfo  {journal} {IEEE
  Transactions on Aerospace and Electronic Systems}\ }\textbf {\bibinfo
  {volume} {57}},\ \bibinfo {pages} {3520--3528} (\bibinfo {year}
  {2021})}\BibitemShut {NoStop}%
\bibitem [{\citenamefont {Hen}\ and\ \citenamefont
  {Spedalieri}(2016)}]{hen2016quantum}%
  \BibitemOpen
  \bibfield  {author} {\bibinfo {author} {\bibfnamefont {Itay}\ \bibnamefont
  {Hen}}\ and\ \bibinfo {author} {\bibfnamefont {Federico~M.}\ \bibnamefont
  {Spedalieri}},\ }\bibfield  {title} {\enquote {\bibinfo {title} {Quantum
  annealing for constrained optimization},}\ }\href {\doibase
  10.1103/PhysRevApplied.5.034007} {\bibfield  {journal} {\bibinfo  {journal}
  {Phys. Rev. Applied}\ }\textbf {\bibinfo {volume} {5}},\ \bibinfo {pages}
  {034007} (\bibinfo {year} {2016})}\BibitemShut {NoStop}%
\bibitem [{\citenamefont {Hadfield}\ \emph {et~al.}(2019)\citenamefont
  {Hadfield}, \citenamefont {Wang}, \citenamefont {O’Gorman}, \citenamefont
  {Rieffel}, \citenamefont {Venturelli},\ and\ \citenamefont
  {Biswas}}]{hadfield2019quantum}%
  \BibitemOpen
  \bibfield  {author} {\bibinfo {author} {\bibfnamefont {Stuart}\ \bibnamefont
  {Hadfield}}, \bibinfo {author} {\bibfnamefont {Zhihui}\ \bibnamefont {Wang}},
  \bibinfo {author} {\bibfnamefont {Bryan}\ \bibnamefont {O’Gorman}},
  \bibinfo {author} {\bibfnamefont {Eleanor~G.}\ \bibnamefont {Rieffel}},
  \bibinfo {author} {\bibfnamefont {Davide}\ \bibnamefont {Venturelli}}, \ and\
  \bibinfo {author} {\bibfnamefont {Rupak}\ \bibnamefont {Biswas}},\ }\bibfield
   {title} {\enquote {\bibinfo {title} {From the quantum approximate
  optimization algorithm to a quantum alternating operator ansatz},}\ }\href
  {https://www.mdpi.com/1999-4893/12/2/34} {\bibfield  {journal} {\bibinfo
  {journal} {Algorithms}\ }\textbf {\bibinfo {volume} {12}} (\bibinfo {year}
  {2019})}\BibitemShut {NoStop}%
\bibitem [{\citenamefont {Stollenwerk}\ \emph
  {et~al.}(2020{\natexlab{b}})\citenamefont {Stollenwerk}, \citenamefont
  {Hadfield},\ and\ \citenamefont {Wang}}]{stollenwerk2020toward}%
  \BibitemOpen
  \bibfield  {author} {\bibinfo {author} {\bibfnamefont {Tobias}\ \bibnamefont
  {Stollenwerk}}, \bibinfo {author} {\bibfnamefont {Stuart}\ \bibnamefont
  {Hadfield}}, \ and\ \bibinfo {author} {\bibfnamefont {Zhihui}\ \bibnamefont
  {Wang}},\ }\bibfield  {title} {\enquote {\bibinfo {title} {Toward quantum
  gate-model heuristics for real-world planning problems},}\ }\href {\doibase
  10.1109/TQE.2020.3030609} {\bibfield  {journal} {\bibinfo  {journal} {IEEE
  Transactions on Quantum Engineering}\ }\textbf {\bibinfo {volume} {1}},\
  \bibinfo {pages} {1--16} (\bibinfo {year} {2020}{\natexlab{b}})}\BibitemShut
  {NoStop}%
\bibitem [{\citenamefont {Garey}\ and\ \citenamefont
  {Johnson}(1979)}]{Garey1979}%
  \BibitemOpen
  \bibfield  {author} {\bibinfo {author} {\bibfnamefont {Michael~R.}\
  \bibnamefont {Garey}}\ and\ \bibinfo {author} {\bibfnamefont {David~S.}\
  \bibnamefont {Johnson}},\ }\href@noop {} {\emph {\bibinfo {title} {Computers
  and Intractability: A Guide to the Theory of NP-Completeness}}}\ (\bibinfo
  {publisher} {W. H. Freeman \& Co.},\ \bibinfo {address} {USA},\ \bibinfo
  {year} {1979})\BibitemShut {NoStop}%
\bibitem [{\citenamefont {Paulson}\ \emph {et~al.}(2021)\citenamefont
  {Paulson}, \citenamefont {Dellantonio}, \citenamefont {Haase}, \citenamefont
  {Celi}, \citenamefont {Kan}, \citenamefont {Jena}, \citenamefont {Kokail},
  \citenamefont {van Bijnen}, \citenamefont {Jansen}, \citenamefont {Zoller},\
  and\ \citenamefont {Muschik}}]{Paulson2020}%
  \BibitemOpen
  \bibfield  {author} {\bibinfo {author} {\bibfnamefont {Danny}\ \bibnamefont
  {Paulson}}, \bibinfo {author} {\bibfnamefont {Luca}\ \bibnamefont
  {Dellantonio}}, \bibinfo {author} {\bibfnamefont {Jan~F.}\ \bibnamefont
  {Haase}}, \bibinfo {author} {\bibfnamefont {Alessio}\ \bibnamefont {Celi}},
  \bibinfo {author} {\bibfnamefont {Angus}\ \bibnamefont {Kan}}, \bibinfo
  {author} {\bibfnamefont {Andrew}\ \bibnamefont {Jena}}, \bibinfo {author}
  {\bibfnamefont {Christian}\ \bibnamefont {Kokail}}, \bibinfo {author}
  {\bibfnamefont {Rick}\ \bibnamefont {van Bijnen}}, \bibinfo {author}
  {\bibfnamefont {Karl}\ \bibnamefont {Jansen}}, \bibinfo {author}
  {\bibfnamefont {Peter}\ \bibnamefont {Zoller}}, \ and\ \bibinfo {author}
  {\bibfnamefont {Christine~A.}\ \bibnamefont {Muschik}},\ }\bibfield  {title}
  {\enquote {\bibinfo {title} {Simulating 2d effects in lattice gauge theories
  on a quantum computer},}\ }\href {\doibase 10.1103/PRXQuantum.2.030334}
  {\bibfield  {journal} {\bibinfo  {journal} {PRX Quantum}\ }\textbf {\bibinfo
  {volume} {2}},\ \bibinfo {pages} {030334} (\bibinfo {year}
  {2021})}\BibitemShut {NoStop}%
\bibitem [{\citenamefont {Avkhadiev}\ \emph {et~al.}(2020)\citenamefont
  {Avkhadiev}, \citenamefont {Shanahan},\ and\ \citenamefont
  {Young}}]{Avkhadiev2020}%
  \BibitemOpen
  \bibfield  {author} {\bibinfo {author} {\bibfnamefont {A.}~\bibnamefont
  {Avkhadiev}}, \bibinfo {author} {\bibfnamefont {P.{\hspace{0.167em}}E.}\
  \bibnamefont {Shanahan}}, \ and\ \bibinfo {author} {\bibfnamefont
  {R.{\hspace{0.167em}}D.}\ \bibnamefont {Young}},\ }\bibfield  {title}
  {\enquote {\bibinfo {title} {Accelerating lattice quantum field theory
  calculations via interpolator optimization using noisy intermediate-scale
  quantum computing},}\ }\href {\doibase 10.1103/physrevlett.124.080501}
  {\bibfield  {journal} {\bibinfo  {journal} {Phys. Rev. Lett.}\ }\textbf
  {\bibinfo {volume} {124}},\ \bibinfo {pages} {080501} (\bibinfo {year}
  {2020})}\BibitemShut {NoStop}%
\bibitem [{\citenamefont {Mazzola}\ \emph {et~al.}(2021)\citenamefont
  {Mazzola}, \citenamefont {Mathis}, \citenamefont {Mazzola},\ and\
  \citenamefont {Tavernelli}}]{Mazzola2021}%
  \BibitemOpen
  \bibfield  {author} {\bibinfo {author} {\bibfnamefont {Giulia}\ \bibnamefont
  {Mazzola}}, \bibinfo {author} {\bibfnamefont {Simon~V.}\ \bibnamefont
  {Mathis}}, \bibinfo {author} {\bibfnamefont {Guglielmo}\ \bibnamefont
  {Mazzola}}, \ and\ \bibinfo {author} {\bibfnamefont {Ivano}\ \bibnamefont
  {Tavernelli}},\ }\bibfield  {title} {\enquote {\bibinfo {title}
  {Gauge-invariant quantum circuits for $u$(1) and yang-mills lattice gauge
  theories},}\ }\href {\doibase 10.1103/PhysRevResearch.3.043209} {\bibfield
  {journal} {\bibinfo  {journal} {Phys. Rev. Research}\ }\textbf {\bibinfo
  {volume} {3}},\ \bibinfo {pages} {043209} (\bibinfo {year}
  {2021})}\BibitemShut {NoStop}%
\bibitem [{\citenamefont {Tilly}\ \emph {et~al.}(2022)\citenamefont {Tilly},
  \citenamefont {Chen}, \citenamefont {Cao}, \citenamefont {Picozzi},
  \citenamefont {Setia}, \citenamefont {Li}, \citenamefont {Grant},
  \citenamefont {Wossnig}, \citenamefont {Rungger}, \citenamefont {Booth},\
  and\ \citenamefont {Tennyson}}]{Tilly2021}%
  \BibitemOpen
  \bibfield  {author} {\bibinfo {author} {\bibfnamefont {Jules}\ \bibnamefont
  {Tilly}}, \bibinfo {author} {\bibfnamefont {Hongxiang}\ \bibnamefont {Chen}},
  \bibinfo {author} {\bibfnamefont {Shuxiang}\ \bibnamefont {Cao}}, \bibinfo
  {author} {\bibfnamefont {Dario}\ \bibnamefont {Picozzi}}, \bibinfo {author}
  {\bibfnamefont {Kanav}\ \bibnamefont {Setia}}, \bibinfo {author}
  {\bibfnamefont {Ying}\ \bibnamefont {Li}}, \bibinfo {author} {\bibfnamefont
  {Edward}\ \bibnamefont {Grant}}, \bibinfo {author} {\bibfnamefont {Leonard}\
  \bibnamefont {Wossnig}}, \bibinfo {author} {\bibfnamefont {Ivan}\
  \bibnamefont {Rungger}}, \bibinfo {author} {\bibfnamefont {George~H.}\
  \bibnamefont {Booth}}, \ and\ \bibinfo {author} {\bibfnamefont {Jonathan}\
  \bibnamefont {Tennyson}},\ }\bibfield  {title} {\enquote {\bibinfo {title}
  {The variational quantum eigensolver: A review of methods and best
  practices},}\ }\href {\doibase https://doi.org/10.1016/j.physrep.2022.08.003}
  {\bibfield  {journal} {\bibinfo  {journal} {Phys. Rep.}\ }\textbf {\bibinfo
  {volume} {986}},\ \bibinfo {pages} {1--128} (\bibinfo {year} {2022})},\
  \bibinfo {note} {the Variational Quantum Eigensolver: a review of methods and
  best practices}\BibitemShut {NoStop}%
\bibitem [{\citenamefont {Amaro}\ \emph {et~al.}(2022)\citenamefont {Amaro},
  \citenamefont {Rosenkranz}, \citenamefont {Fitzpatrick}, \citenamefont
  {Hirano},\ and\ \citenamefont {Fiorentini}}]{Amaro_2022_JSP}%
  \BibitemOpen
  \bibfield  {author} {\bibinfo {author} {\bibfnamefont {David}\ \bibnamefont
  {Amaro}}, \bibinfo {author} {\bibfnamefont {Matthias}\ \bibnamefont
  {Rosenkranz}}, \bibinfo {author} {\bibfnamefont {Nathan}\ \bibnamefont
  {Fitzpatrick}}, \bibinfo {author} {\bibfnamefont {Koji}\ \bibnamefont
  {Hirano}}, \ and\ \bibinfo {author} {\bibfnamefont {Mattia}\ \bibnamefont
  {Fiorentini}},\ }\bibfield  {title} {\enquote {\bibinfo {title} {A case study
  of variational quantum algorithms for a job shop scheduling problem},}\
  }\href {\doibase 10.1140/epjqt/s40507-022-00123-4} {\bibfield  {journal}
  {\bibinfo  {journal} {EPJ Quantum Technology}\ }\textbf {\bibinfo {volume}
  {9}},\ \bibinfo {pages} {5} (\bibinfo {year} {2022})}\BibitemShut {NoStop}%
\bibitem [{\citenamefont {Nannicini}(2019)}]{Nannicini_2019_IBM}%
  \BibitemOpen
  \bibfield  {author} {\bibinfo {author} {\bibfnamefont {Giacomo}\ \bibnamefont
  {Nannicini}},\ }\bibfield  {title} {\enquote {\bibinfo {title} {Performance
  of hybrid quantum-classical variational heuristics for combinatorial
  optimization},}\ }\href {\doibase 10.1103/PhysRevE.99.013304} {\bibfield
  {journal} {\bibinfo  {journal} {Phys. Rev. E}\ }\textbf {\bibinfo {volume}
  {99}},\ \bibinfo {pages} {013304} (\bibinfo {year} {2019})}\BibitemShut
  {NoStop}%
\bibitem [{\citenamefont {Mugel}\ \emph {et~al.}(2022)\citenamefont {Mugel},
  \citenamefont {Kuchkovsky}, \citenamefont {S\'anchez}, \citenamefont
  {Fern\'andez-Lorenzo}, \citenamefont {Luis-Hita}, \citenamefont {Lizaso},\
  and\ \citenamefont {Or\'us}}]{Mugel2022}%
  \BibitemOpen
  \bibfield  {author} {\bibinfo {author} {\bibfnamefont {Samuel}\ \bibnamefont
  {Mugel}}, \bibinfo {author} {\bibfnamefont {Carlos}\ \bibnamefont
  {Kuchkovsky}}, \bibinfo {author} {\bibfnamefont {Escol\'astico}\ \bibnamefont
  {S\'anchez}}, \bibinfo {author} {\bibfnamefont {Samuel}\ \bibnamefont
  {Fern\'andez-Lorenzo}}, \bibinfo {author} {\bibfnamefont {Jorge}\
  \bibnamefont {Luis-Hita}}, \bibinfo {author} {\bibfnamefont {Enrique}\
  \bibnamefont {Lizaso}}, \ and\ \bibinfo {author} {\bibfnamefont {Rom\'an}\
  \bibnamefont {Or\'us}},\ }\bibfield  {title} {\enquote {\bibinfo {title}
  {Dynamic portfolio optimization with real datasets using quantum processors
  and quantum-inspired tensor networks},}\ }\href {\doibase
  10.1103/PhysRevResearch.4.013006} {\bibfield  {journal} {\bibinfo  {journal}
  {Phys. Rev. Research}\ }\textbf {\bibinfo {volume} {4}},\ \bibinfo {pages}
  {013006} (\bibinfo {year} {2022})}\BibitemShut {NoStop}%
\bibitem [{\citenamefont {tA-v~et al.}(2021)}]{Qiskit}%
  \BibitemOpen
  \bibfield  {author} {\bibinfo {author} {\bibfnamefont {A}~\bibnamefont
  {tA-v~et al.}},\ }\href {\doibase 10.5281/zenodo.2573505} {\enquote {\bibinfo
  {title} {Qiskit: An open-source framework for quantum computing},}\ }
  (\bibinfo {year} {2021})\BibitemShut {NoStop}%
\bibitem [{\citenamefont {Powell}(1994)}]{Powell1994}%
  \BibitemOpen
  \bibfield  {author} {\bibinfo {author} {\bibfnamefont {Michael~JD}\
  \bibnamefont {Powell}},\ }\bibfield  {title} {\enquote {\bibinfo {title} {A
  direct search optimization method that models the objective and constraint
  functions by linear interpolation},}\ }in\ \href@noop {} {\emph {\bibinfo
  {booktitle} {Advances in optimization and numerical analysis}}}\ (\bibinfo
  {publisher} {Springer},\ \bibinfo {year} {1994})\ pp.\ \bibinfo {pages}
  {51--67}\BibitemShut {NoStop}%
\bibitem [{Note1()}]{Note1}%
  \BibitemOpen
  \bibinfo {note} {Note that this is possible because we assume a perfect
  quantum computer without shot noise, i.e. we have direct access to the state
  vector, and we can monitor the overlap with exact solution throughout the
  optimization procedure.}\BibitemShut {Stop}%
\end{thebibliography}%
\end{document}